\documentclass[a4paper,11pt]{article}
\pdfoutput=1
\usepackage{jcappub}
\usepackage{journal_abbrev}
\usepackage[svgnames]{xcolor}
\usepackage{xparse, xspace, verbatim, ulem, subcaption, float, comment, bm, bbold, xstring, placeins, graphicx, amsmath, amssymb, soul, hyperref, adjustbox, siunitx, lmodern, enumitem, csquotes}
\usepackage[skip=9pt]{caption}
\xspaceaddexceptions{]\}}

\allowdisplaybreaks

\newcommand{\borg}{{\textsc{borg}}\xspace}

\newcommand{\cic}{\textsc{cic}}
\newcommand{\tophat}{\textsc{th}}

\newcommand{\refeq}[1]{Eq.~(\ref{eq:#1})}          
\newcommand{\refeqs}[2]{Eqs.~(\ref{eq:#1})---(\ref{eq:#2})}          
                   
\newcommand{\reffig}[1]{Figure~\ref{fig:#1}}

\newcommand{\refsec}[1]{Section~\ref{sec:#1}}          
\newcommand{\refapp}[1]{Appendix~\ref{app:#1}}
\newcommand{\reftab}[1]{Table~\ref{tab:#1}}

\def\be{\begin{equation}}
\def\ee{\end{equation}}
\def\bea{\begin{eqnarray}}
\def\eea{\end{eqnarray}}

\def\ba#1\ea{\begin{align}#1\end{align}}
\def\bab#1\eab{\begin{equation}\begin{aligned}[b]#1\end{aligned}\end{equation}}

\def\bg#1\eg{\begin{gather}#1\end{gather}}
\newcommand{\specialcell}[2][c]{%
  \begin{tabular}[#1]{@{}c@{}}#2\end{tabular}}

\newcommand\Msunh{h^{-1}M_{\odot}}
\newcommand\Mpch{\,h^{-1}\mathrm{Mpc}}
\newcommand\iMpch{\,h\,\mathrm{Mpc}^{-1}}

\renewcommand{\lg}{\log_{10}}

\newcommand{\eps}{\epsilon}

\newcommand{\ini}{\mathrm{ini}}
\newcommand{\fwd}{\mathrm{fwd}}
\newcommand{\lin}{\mathrm{lin}}

\newcommand{\rec}{\mathrm{infer}}
\newcommand{\true}{\mathrm{true}}

\def\<{\left\langle}
\def\>{\right\rangle}

\def\A{\mathcal{A}}

\def\C{\mathrm{C}}

\def\N{\mathcal{N}}
\def\O{\mathcal{O}}
\def\P{\mathcal{P}}

\def\R{\mathcal{R}}

\newcommand{\vbf}[1]{\bm{#1}}

\newcommand{\vx}{\vbf{x}}

\newcommand{\vk}{\vbf{k}}

\renewcommand{\L}{\Lambda}

\newcommand{\Om}{\Omega_\mathrm{m}}

\newcommand{\se}{\sigma_\mathrm{8}}
\newcommand{\ns}{n_{\mathrm{s}}}
\def\d{\delta}

\def\ngrid{N_{\mathrm{grid}}}

\newcommand{\avnh}{\overline{n}_h}

\newcommand{\dinput}{\d_\mathrm{input}} 
\newcommand{\drec}{\d_\rec} 
\newcommand{\rri}{r_{\mathrm{ri}}} 
\newcommand{\rhi}{r_{\mathrm{hi}}} 
\newcommand{\dm}{\d_m} 
\renewcommand{\dh}{\d_h} 
\newcommand{\nh}{n_h}
\newcommand{\nhdet}{n_{h,\mathrm{det}}}
\newcommand{\dhdet}{\d_{h,\mathrm{det}}} 

\def\boxone{\text{Box}-1\xspace}
\def\boxtwo{\text{Box}-2\xspace}
\def\Lbox{L_\text{box}}
\def\lgrid{l_\text{grid}}

\def\Plin{P_\text{L}}

\def\emph#1{\textit{#1}}

\newcommand{\samelinetwofig}[4]{%
     \begin{subfigure}[b]{0.49\textwidth}
         \centering
         \includegraphics[width=\textwidth]{#1}
     \end{subfigure}
     \hfill
     \begin{subfigure}[b]{0.49\textwidth}
         \centering
         \includegraphics[width=\textwidth]{#2}
     \end{subfigure}
     \caption{#4}
     \label{#3}
}

\newcommand{\twolinefourfig}[6]{%
     \begin{subfigure}[b]{0.49\textwidth}
         \centering
         \includegraphics[width=\textwidth]{#1}
     \end{subfigure}
     \hfill
     \begin{subfigure}[b]{0.49\textwidth}
         \centering
         \includegraphics[width=\textwidth]{#2}
   	 \end{subfigure}\vspace{-1em}
       
     \begin{subfigure}[b]{0.49\textwidth}
         \centering
         \includegraphics[width=\textwidth]{#3}
     \end{subfigure}
     \hfill
     \begin{subfigure}[b]{0.49\textwidth}
         \centering
         \includegraphics[width=\textwidth]{#4}
     \end{subfigure}     
     \caption{#6}
     \label{#5}
}

\title{Impacts of the physical data model on the \emph{forward} inference of initial conditions from biased tracers}

\author[a]{Nhat-Minh Nguyen,}
\author[a]{Fabian Schmidt,}
\author[b]{Guilhem Lavaux,}
\author[c]{and Jens Jasche}

\emailAdd{minh@mpa-garching.mpg.de}
\emailAdd{fabians@mpa-garching.mpg.de}
\emailAdd{guilhem.lavaux@iap.fr}
\emailAdd{jens.jasche@fysik.su.se}

\affiliation[a]{Max-Planck-Institut f\"ur Astrophysik, Karl-Schwarzschild-Str. 1, 85748 Garching, Germany}
\affiliation[b]{Sorbonne Universit\'{e}, CNRS, UMR 7095, Institut d'Astrophysique de Paris, 98 bis bd Arago, 75014 Paris, France}
\affiliation[c]{Department of Physics, Stockholm University, Albanova University Centre, SE-106 91 Stockholm, Sweden}

\abstract{
  We investigate the impact of each ingredient in the employed physical data model on the Bayesian forward inference of initial conditions from biased tracers at the field level. Specifically, we use dark matter halos in a given cosmological simulation volume as tracers of the underlying matter density field. We study the effect of tracer density, grid resolution, gravity model, bias model and likelihood on the inferred initial conditions. We find that the cross-correlation coefficient between true and inferred phases reacts weakly to all ingredients above, and is well predicted by the theoretical expectation derived from a Gaussian model on a broad range of scales. The bias in the amplitude of the inferred initial conditions, on the other hand, depends strongly on the bias model and the likelihood.
  We conclude that the bias model and likelihood hold the key to an unbiased cosmological inference. Together they must keep the systematics---which arise from the sub-grid physics that are marginalized over---under control in order to obtain an unbiased inference.}
\keywords{initial conditions from LSS, forward modeling, field-level statistics, cosmological inference, redshift surveys, dark matter halos, bias}
\begin{document}

\maketitle
\flushbottom

\section{Introduction}
\label{sec:intro}

The distribution of large-scale structures (LSS) in our Universe, specifically galaxies---biased tracers of the underlying matter density field---is very far from the random distribution of initial quantum fluctuations that seed their formation. As complex as the process of galaxy formation is, on large and quasi-linear scales, the total matter distribution follows that of dark matter (DM) which only feels the effect of gravity. Going forward in time, the gravitational evolution of matter distribution is fully \emph{deterministic}. It follows that, by forward-modeling this process, together with the relation between matter and galaxy distributions, one should be able to retrieve a vast amount of information directly from the three-dimensional distributions of galaxies as probed by galaxy redshift surveys. This includes, but is not limited to the initial conditions of our observed patch of the Universe and parameters in our cosmological models.

Standard approaches in cosmological inference and reconstruction from LSS data, however, often rely on either modeling of $n$-point functions of galaxies \citep[e.g.][]{Gil-Marin:2017, Alam:2017, Ivanov:2020, DES:2020} or reversing the evolution of matter fluctuations by moving mass particles or galaxies backward (reconstruction hereafter) \citep[e.g.][]{Frisch:2002, Padmanabhan:2012, Schmittfull:2017, Wang:2020}. Here, we are interested in the aforementioned more direct route: \emph{forward modeling the matter and galaxy clustering directly at the field-level} \cite{Jasche:2013, Wang:2014, Jasche:2015, Ata:2015, Wang:2016, Ata:2017, Modi:2018, Jasche:2019, Lavaux:2019, Schmidt:2019, Cabass:2019b, Cabass:2020a, Cabass:2020b, Elsner:2020, Schmidt:2020a, Schmidt:2020b}.
This alternative path not only circumvents the difficult and cumbersome problem of how to accurately predict and compute the higher $n$-point functions (plus their covariances) from theory \cite{Takada:2003, Sefusatti:2006, Lazanu:2017, Hashimoto:2017, Sugiyama:2020} and data \cite{Gil-Marin:2017, Slepian:2017}, respectively, but also, in principle, accesses all information available in the three-dimensional data field above the smoothing scale, regardless of its Gaussianity.
In fact, the latter statement is not guaranteed to apply to the $n$-point function approach if the observed galaxy field is non-Gaussian at the scales of interest \cite{Carron:2011}.

\begin{figure*}[htbp]%
  \includegraphics[width=\textwidth]{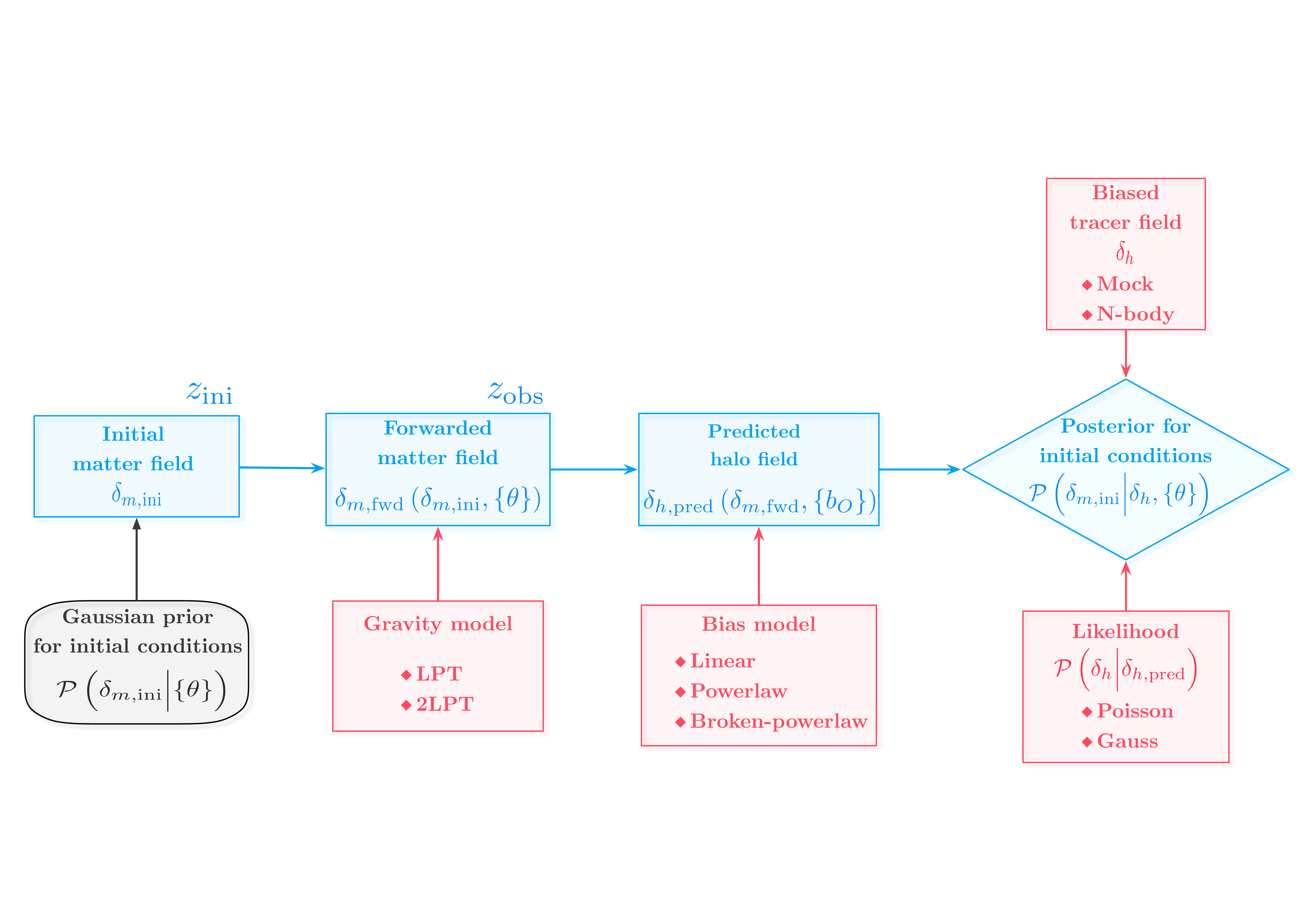}
  \protect\caption{Simplified flowchart of the forward inference framework, as relevant for this work. The input data is illustrated by the top red panel. The target of the inference is represented by the blue diamond panel in the middle row. Different ingredients of the physical data model are displayed in the bottom row. We highlight the ingredients inspected in this work in red. Different options for each inspected ingredient are explicitly listed in the corresponding panels. Our only anchored assumption is the Gaussian prior on initial conditions, shown in the round-edge gray panel.
  \label{fig:flowchart}}
\end{figure*}

There are four principal ingredients of the physical data model in such a field-level, forward inference approach \cite{Jasche:2013, Schmidt:2019}:
\begin{enumerate}[leftmargin=1.5cm]%
	\item \textbf{Prior for initial conditions}: This element encodes our prior knowledge of the statistical distribution of matter density fluctuations at the very early stage of the Universe that comes from the theory of initial fluctuations \cite{BBKS:1986, Weinberg:2008} and observation of the Cosmic Microwave Background (CMB) \cite{WMAP:2007, Planck:2018IX}.
	\item \textbf{Gravity model for matter}: This ingredient evolves matter distributions forward in time, from a specific set of initial conditions, under the effect of gravity.
	\item \textbf{Bias model for deterministic matter-galaxy bias}: This constituent relates the forward-evolved matter field and the deterministic galaxy field by means of a deterministic relation (see \cite{Desjacques:2018} for a review).
	\item \textbf{Likelihood for stochastic galaxy bias}: This final piece attempts to model the scatter in the deterministic matter-galaxy bias relation, i.e. the difference between the deterministic and observed galaxy fields, in the form of a conditional probability \cite{Schmidt:2019, Cabass:2019b, Cabass:2020a, Cabass:2020b}.
\end{enumerate}
We summarize the connection between these ingredients of the physical data model in \reffig{flowchart}, where (and hereafter) we have generalized \enquote{galaxy} to \enquote{biased tracer}, partially due to the fact that our results are obtained using DM halos in N-body simulations, and also because most of our results and arguments apply to all generic biased tracers. We will discuss possible exceptions at the end of \refsec{results_likelihood}.

How strongly does the quality of a forward inference depend on these ingredients? In particular, how sensitive are the inferred initial conditions to different choices for each ingredient? So far, this is an open question in the context of forward modeling (see \cite{Birkin:2019} for a case study in the \enquote{backward} modeling approach).
It is the aim of this paper to address that question. Specifically, inspired by previous findings in \cite{Elsner:2020, Schmidt:2020a, Schmidt:2020b}, our main goal is to \emph{identify the key ingredients for a robust, unbiased cosmological forward inference from LSS}, and what should be the focus of future studies to realize this goal.

In this paper, we specifically measure: the cross-correlation coefficient between true and inferred phases (phase-correlation hereafter) $r(k)$ and the amplitude-bias in the corresponding power spectra (which is often referred to by the literature of reconstruction, and hence hereafter as \enquote{transfer function}) $T_{\rec}(k)$.
Clearly, the former is important for studies making use of the inference for cross-correlations with other probes.
The latter is important as it indicates systematic biases in cosmological parameter that are to be expected.
Specifically, a  transfer function that is biased in a \emph{scale-independent} way implies a bias in the inferred primordial normalization of the power spectrum $\A_s$, whereas a \emph{scale-dependently biased} transfer function entails biases in not only $\A_s$ but other cosmological parameters, e.g. $\Om$, $\ns$, as well.

We use DM main halos identified in N-body simulations, for which the true initial conditions are known, as physical biased tracers.
We further restrict our analysis to the halo rest frame, i.e. the case without redshift space distortion (RSD).
We consider multiple inference setups corresponding to different choices for the ingredients two to four in the above list (bottom red panels in \reffig{flowchart}), as well as different options for halo number density and grid resolution, i.e. smoothing scale.

All forward inferences in this paper follow exactly the flowchart in \reffig{flowchart}, and are carried out using the \borg algorithm that was first introduced in \cite{Jasche:2013}.
As our parameter space actually consists of the whole three-dimensional field of initial conditions, augmented by the bias parameters, efficiently Monte-Carlo Markov-chain (MCMC) sampling this very high-dimensional space of the posterior is a truly daunting task.
The task is even more critical in the particular problem of forward inference, for each sample requires essentially a  full simulation from the early-time Gaussian initial conditions to the late-time non-Gaussian observed data, as diagramed in \reffig{flowchart}, hence some non-negligible numerical expense.
The \borg algorithm addresses this issue by utilizing the Hamiltonian Monte Carlo (HMC) sampling method \cite{Duane:1987, Neal:2012, Jasche:2010b} which combines geometrical information of the posterior, encoded in its gradient, and Hamiltonian dynamics. The latter ensures that the Metropolis-Hastings proposal always has a high acceptance rate---providing that errors in the numerical integration of the Hamiltonian equation of motion are kept under control \cite{Neal:2012, Betancourt:2014}.
We encourage readers to see \cite{Jasche:2013} for further conceptual and technical details specifically to \borg. We note that, in an application on real data from a galaxy redshift survey, \borg is also capable of routinely dealing with multiple sources of systematics, including survey geometries, selection functions, foreground contaminations and RSD effects \citep[See, for example][]{Lavaux:2019}. These steps are not included in \reffig{flowchart} as they are not relevant for our analysis using rest-frame halos in N-body simulations.

Our results in \refsec{results} are based on our analyses of the outputs of multiple \borg runs: each is an ensemble of MCMC samples that resembles the corresponding posterior density distribution. In our particular case, after marginalizing over bias parameters, the posterior parameter space is the three-dimensional field of initial conditions (middle blue diamond panel in \reffig{flowchart}), physically constrained by the three-dimensional input field of halos (top red square panel in \reffig{flowchart}) and our choices of models for each ingredient (bottom red rectangle panels).

The rest of this paper is organized as follows.
In \refsec{setup}, we describe in more details the three ingredients under investigation and other settings that together characterize the configuration of each inference. We further detail the N-body simulations and the halos that are used as the input data in \refsec{tracers}. In \refsec{results}, we first define the estimators used to assess the quality of the inference and later present our results. We then highlight the ingredients that demand better modeling in \refsec{discussion}, and conclude in \refsec{conclusion}. The appendices provide additional information about the derivation of the Gaussian expectation in \refsec{results}, numerical implementation details and bias priors relevant for \refsec{result_GADGET2}, as well as MCMC chains analyzed throughout this paper.

\section*{Convention and notation}
\label{sec:notation}

Our Fourier notation follows that of \cite{Desjacques:2018, Schmidt:2019}:
\bab
f(\vk) &\equiv \int d^3\vx f(\vx) e^{-i\vk\cdot\vx} \equiv \int_{\vx} f(\vx)e^{-i\vk\cdot\vx} \\
f(\vx) &\equiv \int \frac{d^3\vk}{(2\pi)^3} f(\vk) e^{i\vk\cdot\vx} \equiv \int_{\vk}f(\vk)e^{i\vk\cdot\vx}.
\label{eq:Fourier_notation}
\eab

For ensemble averages, we distinguish between $\<O\>_k$, which denotes an average of $O$ over Fourier wavenumber $k$, and $\<O\>_s$, which denotes an average of $O$ over MCMC samples. A spatial-average of $O$, on the other hand, is denoted by $\bar{O}$.

The matter and halo density contrast fields (on a fixed time slice) are defined as
\be
\d_{m}(\vx) \equiv \frac{\rho_m(\vx)}{\bar{\rho}_m}-1,\quad \d_{h}(\vx) \equiv \frac{n_{h}(\vx)}{\bar{n}_{h}}-1.
\label{eq:density_contrast}
\ee

Throughout the paper, we assume adiabatic, growing-mode initial conditions such that the initial fluctuations in the matter field can be described by only a single field $\d_{m,\ini}$.
Further, we adopt a flat $\Lambda$CDM cosmology with a primordial matter power spectrum of the form \cite{Dodelson:2021}:
\be
P_{\R}(k) = 2\pi^2\A_s k^{-3}\left(\frac{k}{k_{\mathrm{pivot}}}\right)^{n_s-1},
\label{eq:primordial_Pk}
\ee
such that the linear matter power spectrum is given by \cite{Dodelson:2021}:
\be
\Plin(k,a,\theta) = \frac{8\pi^2}{25}\frac{\A_s}{\Om^2}D_+^2(a)T^2(k)\frac{k^\ns}{H_0^4k_{\mathrm{pivot}}^{\ns-1}},
\label{eq:linear_Pk}
\ee
where $\theta$ represents the set of cosmological parameters $\left(\Om,\A_s,\ns,H_0\right)$, namely the total matter density parameter, the primordial spectrum normalization, the scalar spectral index and the Hubble constant, respectively; while $T(k)$ denotes the matter transfer function.
We also adopt a pivot scale of $k_{\mathrm{pivot}}=0.05\Mpch$ following the Planck team's convention \cite{Planck:2018VI}.
We further assume the prior on the distribution of $\d_{m,\ini}\equiv\left(\d_{m,\ini}(\vx_1),...,\d_{m,\ini}(\vx_n)\right)$, i.e. ingredient one in the above list, to follow a zero-mean, multivariate Gaussian
\be
\P_{\mathrm{prior}} \left(\d_{m,\ini} \Big | \theta\right) = \N \left(\d_{m,\ini} \Big | \mu=0, \C=\<\d_{m,\ini}(\vx_i)\d_{m,\ini}(\vx_j)\>\right).
\label{eq:multivariate_Gaussian_ICs}
\ee
In a Fourier-space representation of $\d_{m,\ini}$, the covariance matrix $\C$ is diagonal, and the diagonal is set by the linear matter power spectrum in \refeq{linear_Pk}, at $a=a_{\ini}$.

For all inferences presented in this work, we proceed by fixing all cosmological parameters to the values used in the simulations, \emph{with one exception}. Instead of specifying $\A_s$, we specify the normalization of the present linear matter power spectrum $\se$, which is defined as
\be
\se^2 = \sigma^2\left(R_{\tophat}=8\Mpch, z=0\right) = \int_{\vk}\, \Plin\left(k, z=0\right)\,W^2_{\tophat}\left(R_{\tophat}=8\Mpch, k\right)
\label{eq:se}
\ee
 where $W_{\tophat}(R_{\tophat},k)$ is the Fourier transform of the top-hat filter, and fix it to the value in the corresponding simulation. This is due to the convention in the current numerical implementation of \borg. In practice, this subtlety might lead to a percent-level bias in the measured transfer function $T_{\rec}(k)$, especially if the forward inference and the input simulation employ different matter transfer functions $T(k)$. In our case, \borg adopts the Eisenstein-Hu fitting formula \cite{eisenstein/hu:1998, eisenstein/hu:1999} while our N-body simulations take the matter transfer functions computed with CAMB or CLASS as inputs (cf. \refsec{tracers}). The implication of this difference is examined in more details in \refapp{normalization_bias}. In fact, the systematic shift this effect might induce on $T_{\rec}(k)$ is negligible compared to the bias we observed throughout \refsec{result_GADGET2}.

The numerical implementation of gravity and bias models in the \borg framework discretizes the matter and halo density fields using a cloud-in-cell (CIC) projection onto a grid with cell size $\lgrid$ such that
\be
\d_{h,m}^i \equiv \int d^3x\, W_{\cic}(\lgrid, \vx, \vx^i)\, \d_{h,m}(\vx^i-\vx)
\label{eq:CIC_projection}
\ee
where the superscript $i$ denotes the cell index and $W_{\cic}(\vx, \vx^i)$ is the real-space, normalized CIC filter:

\begin{subequations}
\begin{align}
W_{\cic}(\lgrid, \vx, \vx^i)&=\frac{1}{\lgrid}\,\begin{cases}
1-|\vx^i-\vx|/\lgrid &\text{$|\vx^i-\vx|<\lgrid$} \\
0 &\text{otherwise},
\end{cases}
\label{eq:CIC_filter_realspace}
\intertext{whose Fourier-space representation is given by}
W_{\cic}(\lgrid, k) &=\mathrm{sinc}^2\left(\frac{k\,\lgrid}{2}\right).
\label{eq:CIC_filter_fourierspace}
\end{align}
\end{subequations}

Let us briefly discuss implications of the shape of the smoothing filter here. A sharp-$k$ filter, for example, corresponds to a particular scale cut-off $\Lambda$. This means \emph{only modes below the cut-off scale $\Lambda$ enter the analysis}, and the higher this cut-off is, the more non-linear modes are included. The CIC filter adopted here, on the other hand, is not localized in Fourier space and thus \emph{does not completely remove modes with $k > 1/\lgrid$}. This might allow non-linear modes to slip into the inference, as further discussed in \refsec{results_bias_model} and \refsec{results_likelihood}.

\section{Inference setup}
\label{sec:setup}

Below, for each inspected ingredient in the inference, we introduce and discuss each model or option considered within this study.

\subsection{Gravity model}
\label{sec:ICborg:forward}

We choose the first- and second-order Lagrangian perturbation theories (LPT and 2LPT) \citep[e.g.][]{Moutarde:1991, Bouchet:1992, Bouchet:1995} as our gravity models for this comparison.
We refer readers to \cite{Jasche:2013} for specific details about the numerical implementation of these models in the \borg framework employed here.

Our two choices are the most common and relevant for forward inference of current galaxy survey data \citep[see, for example][]{Lavaux:2019}.
A forward inference powered by any of the two approximations naturally includes optimal Baryon Acoustic Oscillation (BAO) reconstruction (to the degree that the forward model is accurate). Especially, with 2LPT, the final joint posterior should also include sub-leading effects of large-scale perturbation modes  For a more detailed discussion on the relation with BAO reconstruction, see Section 6 of \cite{Schmidt:2019}.

\subsection{Bias model}
\label{sec:biasmodel}

In this paper, we consider one linear and two non-linear, non-perturbative bias models. All are Eulerian, local-in-matter-density-field (LIMD) \cite{Desjacques:2018} bias relations, i.e. the biased tracer density at one location only depends on the matter density at that same location in the Eulerian frame.
These models are the ones currently implemented in \borg and recently employed for the analyses in \cite{Lavaux:2016, Jasche:2019, Lavaux:2019, Ramanah:2019, Wang:2020}.

Our actual observable is the tracer count at a given cell, $n_h^i$.
The deterministic or mean-field prediction for this quantity, $\nhdet^i$, 
is related to the quantity $\dhdet^i$ returned by a bias model for the same cell through a nuisance parameter $n_0$ such that
\be
\nhdet^i\equiv n_0\left(1+\dhdet^i\right).
\label{eq:n0}
\ee
Note that, in practice, the inferred value of $n_0$ can be different from the true mean number of halos in a grid cell $\<\nh^i\>\equiv N_h/(\ngrid^3)$.

\subsubsection{Linear bias}
The linear bias model links the forward-evolved matter density $\d_{m,\fwd}^i$ to the deterministic halo density $\dhdet^i$, via a single bias coefficient $b_1$:
\be
\dhdet^i = b_1\d_{m,\fwd}^i.
\label{eq:linear_biasmodel}
\ee
It is worth noting that the linear bias coefficient $b_1$ in \refeq{linear_biasmodel} is different from the more familiar large-scale linear bias coefficient measured from $n$-point functions \cite{Tinker:2008, Lazeyras:2016}. In fact, all bias coefficients appearing in the field-level approach are \emph{moment biases} (cf. Section 4.2 of \cite{Desjacques:2018}), which explicitly depend on both the specific shape of the smoothing filter and the smoothing scale.
In particular, on linear and quasi-linear scales where the perturbation theory (PT) for matter and biased tracers is guaranteed to converge to the correct result (when carried out to sufficiently high order), the moment bias $b_1^m$ is related to the large-scale $b_1^l$ as follows \cite{Desjacques:2018}
\be
b_1^m(\lgrid) = b_1^l + \O\left(R_\ast^2\sigma_1^2(\lgrid)\right),
\label{eq:moment_large-scale_bias_relation}
\ee
where $R_\ast$ and $\sigma^2_1\equiv \int_{\vk} k^2 \Plin(k) W_{\cic}^2(k)$ are the physically relevant scale for the tracer formation process and the first generalized spectral moment, respectively.
Given the grid resolutions considered in this study (cf. \refsec{grid_resolution}), they yield a leading order correction term of $\O\left(R_\ast^2\sigma_1^2(\lgrid)\right)\simeq0.005-0.053$ and $\simeq0.050-1.327$ for the two N-body simulation setups (cf. \refsec{tracers}).

\subsubsection{Power-law bias}

\begin{figure*}[thbp]
\centerline{\resizebox{0.6\textwidth}{!}{
	\includegraphics*{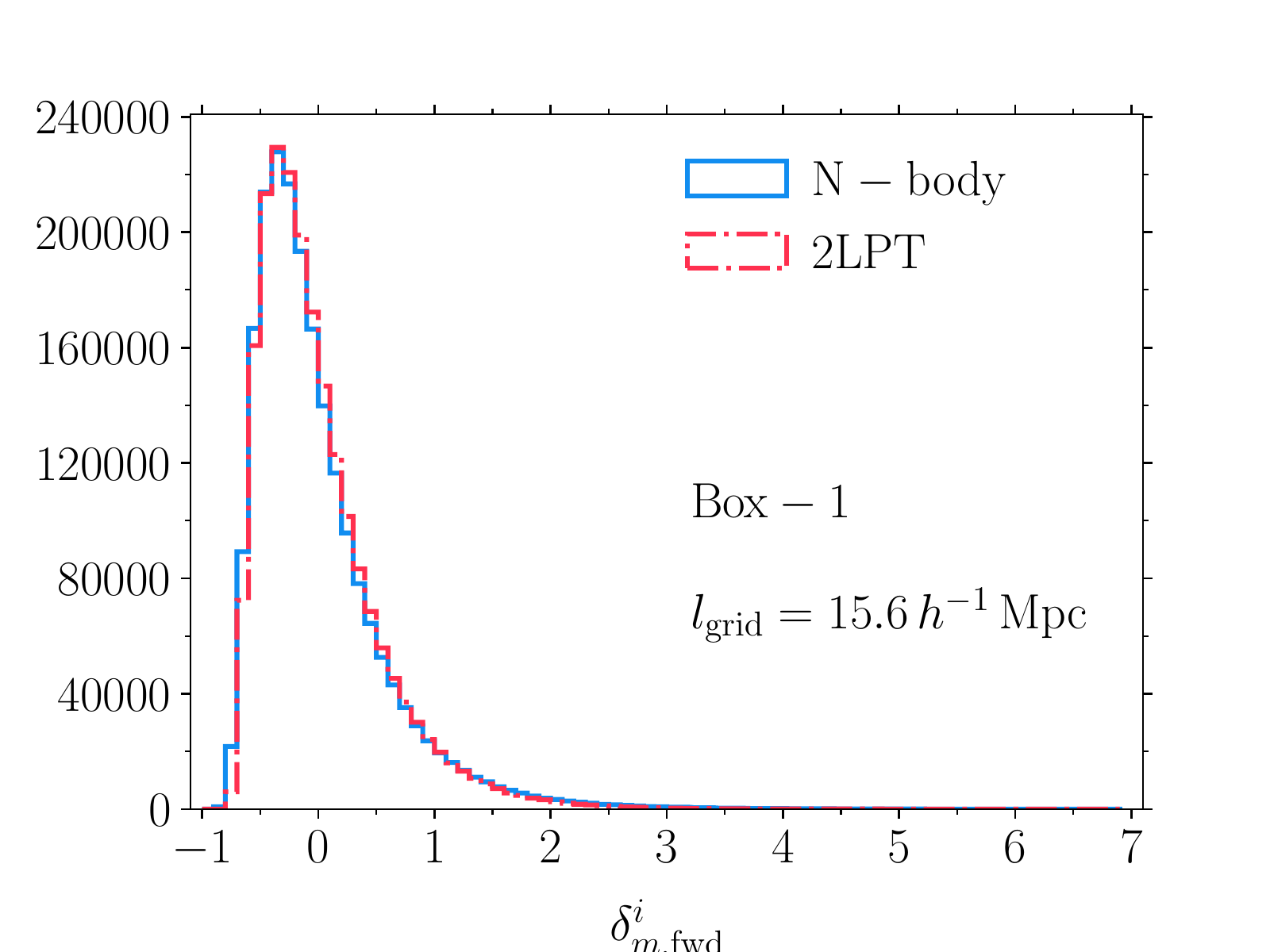}}}\caption{Histograms of the forward-evolved matter density fields $\d^i_{m,\fwd}$ in our fiducial simulation box. The solid blue histogram denotes the field evolved by N-body simulation while the dotted-dashed red histogram represents the one evolved by 2LPT. The total number of grid cells is $\ngrid=128^3$.}	\label{fig:evolved_density_histogram_L2000_N128}
\end{figure*}

In the power-law bias model, the deterministic halo density in a given cell is related to the matter density at that cell also by only one free parameter, the power-law index $\beta$ \cite{Lavaux:2016}
\be
\dhdet^i = \left(1+\d_{m,\fwd}^i\right)^{\beta}-1.
\label{eq:powerlaw_biasmodel}
\ee
For small perturbations, $\d_{m,\fwd}^i\ll1$, \refeq{powerlaw_biasmodel} simplifies to
\be
\dhdet^i = \beta\d_{m,\fwd}^i + \frac{\beta(\beta-1)}{2}\left(\d_{m,\fwd}^i\right)^2 + \O\left(\d_{m,\fwd}^3\right).
\label{eq:powerlaw_biasmodel_approx}
\ee
Thus, for cells where the evolved matter density fluctuations are still small, the role of the power-law index in \refeq{powerlaw_biasmodel} reduces to that of the linear (moment) bias coefficient in \refeq{linear_biasmodel}, i.e. $\beta\simeq b_1$ for $\d_{m,\fwd}^i\ll1$. As shown in \reffig{evolved_density_histogram_L2000_N128}, however, there is a significant fraction of cells where fluctuations, evolved by either N-body or 2LPT, have become highly non-linear for the typical grid resolution we study in this paper. Thus we expect no convergence of \refeq{powerlaw_biasmodel_approx}, but instead significant deviation of $\beta$ from $b_1$ (cf. right panel of \reffig{trace_plot_examples}).
We will return to this point again in the discussion near the end of \refsec{results_likelihood}, where we will see that the same issue poses a problem for the validity of the Poisson likelihood (for halos).

\subsubsection{Broken power-law bias}

The broken power-law bias model, as its name indicates, introduces an additional power-law cut-off into the power-law bias relation in \refeq{powerlaw_biasmodel}, such that
\be
\dhdet^i = \left(1+\d_{m,\fwd}^i\right)^{\beta}\,\exp\left[-\rho(1+\d_{m,\fwd}^i)^{-\eps}\right] - 1.
\label{eq:broken_powerlaw_biasmodel}
\ee
This model essentially behaves like the power-law bias model but accounts for the exponential suppression of halo clustering inside voids \cite{Neyrinck:2014}. It is characterized by three bias parameters: the power-law index $\beta$ and two hyperparameters $(\rho, \eps)$ appearing in the exponential cut-off.

We emphasize again that both the power-law and broken power-law bias are LIMD bias models as the right-hand sides (r.h.s) of \refeq{powerlaw_biasmodel} and \refeq{broken_powerlaw_biasmodel} both include only one single operator constructed out of the local matter density field, that is $\d_{m,\fwd}^i$.

\subsection{Likelihood}
\label{sec:likelihood}

For this case study, we compare Poisson and Gaussian distributions for the likelihood which captures stochasticity in the actual biased tracer field. These two distributions are the most commonly studied and considered in the literature of halo stochasticity \citep[e.g.][]{Miranda:2002, Baldauf:2013}. Together, they form the basis for all active likelihoods in \borg \cite{Lavaux:2016, Jasche:2019, Lavaux:2019, Ramanah:2019}.

The joint likelihood is a product of single-cell likelihoods evaluated at each individual cell, such that
\be
\ln\P\left(n_h \Big|\{\d_{m,\fwd}\}\right) = \sum_{i=1}^{N_g^3} \ln\P^{(1)}\left(\nh^i \Big|\nhdet^i\right)\,,
\label{eq:Pcond}
\ee
where $\P^{(1)}$ is defined as in \refeq{cell_Poisson_likelihood} (for the Poisson likelihood) or \refeq{cell_Gaussian_likelihood} (for the Gaussian likelihood).
The assumption in \refeq{Pcond} is that the stochasticity is uncorrelated between grid cells.

\subsubsection{Poisson likelihood}

The single-cell Poisson likelihood is given by
\be
\P_{\mathrm{Poisson}}^{(1)}\left(\nh^i\Big|\nhdet^i\right) \equiv \frac{\left(\nhdet^i\right)^{\nh^i}\,e^{-\nhdet^i}}{\left(\nh^i\right)!}\,,
\label{eq:cell_Poisson_likelihood}
\ee
where $\nhdet^i$ is defined as in \refeq{n0}.
The Poisson likelihood requires that $\nhdet^i\geq0$.

\subsubsection{Gaussian likelihood}

The single-cell Gaussian likelihood can be written as
\be
\P_{\mathrm{Gaussian}}^{(1)}\left(\nh^i\Big|\nhdet^i\right) \equiv \frac{1}{\sqrt{2\pi}\sigma}\,\exp\left[-\frac{\left(\nh^i-\nhdet^i\right)^2}{2\sigma^2}\right]\,,
\label{eq:cell_Gaussian_likelihood}
\ee
in which the Gaussian standard deviation $\sigma$ is an additional nuisance parameter and assumed to be \emph{universal} for all cells. We will return to this assumption in \refsec{results_likelihood}.

\subsection{Grid resolution}
\label{sec:grid_resolution}

To examine the effect of the smoothing scale, we run our inferences at five different grid resolutions: $\lgrid=31.3,\,20.8,\,15.6$ $\Mpch$ with \boxone (see below) plus its mock counterparts; and $\lgrid=15.6,\,7.8,\,5.2$ $\Mpch$ with \boxtwo plus its mock counterparts. These correspond to a grid of $\ngrid=64^3,\,96^3,\,128^3$ and $\ngrid=32^3,\,64^3,\,96^3$, respectively.

\section{N-body simulation setup}
\label{sec:tracers}

We use one of the GADGET-2 \cite{Springel:2005} N-body simulations first presented in \cite{Biagetti:2016ywx} and later used in \cite{Schmidt:2019, Elsner:2020, Schmidt:2020a} as our main input data. We label this \emph{fiducial box} as \boxone. \boxone has a box size of $\Lbox = 2000\,\Mpch$ and a total number of $N_{\mathrm{part}}=1536^3$ DM particles, giving it a mass resolution of $M_\mathrm{part} = 1.8 \times 10^{11}\,\Msunh$. The simulation adopted a flat $\L$CDM cosmology with $\Om = 0.30$, $\ns = 0.967$, $h = 0.7$, $\se = 0.850$. The input matter transfer function was computed with CLASS\footnote{\url{http://class-code.net}} \cite{Lesgourgues:2011}. Initial conditions for the N-body run were generated at redshift $z_{\ini}= 99$ using the 2LPTic algorithm \cite{Crocce:2006, Scoccimarro:2012}.
DM halos were identified at redshift $z=0$ as spherical over-densities (SO) \cite{Gunn:1972, Press:1974, Kravtsov:2012} using the Amiga Halo Finder (AHF) algorithm\footnote{\url{http://popia.ft.uam.es/AHF/}}
\cite{Gill:2004, Knollmann:2009}, where an over-density threshold of 200 times the background matter density is chosen.
Note that we only include main halos below $M_h=10^{14.5}\Msunh$ in our analysis. We further divide these halos into three mass bins of $\lg M_h=[13.0,13.5),\,[13.5-14.0),\,[14.0-14.5)\,\Msunh$ with the first mass bin being our \emph{fiducial sample}.
      Their corresponding comoving number densities are listed in \reftab{halos}.

Whenever a higher grid resolution inference is required to verify a trend observed in lower resolution inferences, we employ a smaller GADGET-2 simulation box, first presented in \cite{Lazeyras:2016}, as our input data. We label this as \boxtwo. \boxtwo has a box length $\Lbox = 500 \, \Mpch$ and $N_{\mathrm{part}}=512^3$ particles, yielding a mass resolution of $M_\mathrm{part} = 7.0 \times 10^{10} \, \Msunh$.
\boxtwo also assumed a flat $\L$CDM cosmology, albeit with a slightly different cosmology: $\Om = 0.27$, $\ns = 0.950$, $h = 0.7$, $\se = 0.831$.
The input matter transfer function was computed with CAMB\footnote{\url{http://camb.info}} \cite{Lewis:1999}.
Halos were identified in the same fashion as done in \boxone. We consider only a single mass bin of $\lg M_h=[12.55,13.55)\,\Msunh$ in our analysis of \boxtwo.

We summarize the halo samples in \reftab{halos}.
It is worth emphasizing that all inferences take halos in their own rest frame as input, i.e. neglecting the RSD effect.

\begin{table}[b]
\centering
\begin{tabular}{c c c c}
\hline
\hline
Simulation & \specialcell{Mass range\\$\log_{10} M [\Msunh]$} & $N_h$ & $\bar{n}_h\ [(\iMpch)^3]$ \\
\hline
\boxone & $[13.0, 13.5)$ & 2672481 & $3.34\times10^{-4}$ \\
\boxone & $[13.5, 14.0)$ & 882123 & $1.10\times10^{-4}$ \\
\boxone & $[14.0, 14.5)$ & 232654 & $2.91\times10^{-5}$ \\
\hline
\boxtwo & $[12.55, 13.55)$ & 127722 & $1.02\times10^{-3}$ \\
\hline
\hline
\end{tabular}
\caption{The halo samples used in our inferences. Throughout, masses $M \equiv M_{200m}$ are spherical-overdensity masses with respect to 200 times the background matter density.}
\label{tab:halos}
\end{table}

\section{Results}
\label{sec:results}

The phase-correlation coefficient between two given fields $\d_1$, $\d_2$ is given by:
\be
r_{12}(k) = \frac{\<\d_1 \d_2\>_k}{\sqrt{\<\d_1\d_1\>_k\,\<\d_2\d_2\>_k}}
\label{eq:fourier_correlation_general}
\ee
where $\<\,\>_k$ denotes the wavenumber average.
Hereafter, we define the correlation coefficient \refeq{fourier_correlation_general} between the inferred initial modes in the MCMC sample $s'$ and the true ones as
\be
\rri^{s'}(k) = \frac{\<\drec^{s'} \dinput\>_k}{\sqrt{\<\<\drec^s\>_s\<\drec^s\>_s\>_k\,\<\dinput\dinput\>_k}}
\label{eq:fourier_correlation}
\ee
where $\<\,\>_s$ denotes the \borg ensemble average, i.e.
\be
\<O\>_s \equiv \frac{1}{N_{\mathrm{samples}}} \sum_{s}^{N_{\mathrm{samples}}} O^{s},
\ee
where $N_{\mathrm{samples}}=1401$ is the number of MCMC samples considered for each inference in our analysis.
Additionally, we have shortened the notation by writing $\d_{m,\ini}^{\rec}\equiv\drec$ and $\d_{m,\ini}^{\mathrm{input}}\equiv\dinput$. 
Our estimator of the correlation coefficient for each MCMC chain is then the \borg ensemble average of \refeq{fourier_correlation}:
\be
\rri(k) = \<\rri^{s}(k)\>_{s}
\label{eq:rri_mean}
\ee
whose variance is estimated using the sample variance over \borg samples,
\be
\sigma^2_r(k) = \frac{1}{N_{\mathrm{samples}}}\sum_{s'}^{N_{\mathrm{samples}}} \left[\rri^{s'}(k)-\<\rri^s(k)\>_{s}\right]^2 .
\label{eq:rri_var}
\ee
The inferred phases of the initial conditions are unbiased if $\rri(k)$ is consistent with 1 within the uncertainty.
Note that we use the MCMC ensemble mean $\<\<\drec\>_s\>_k$ instead of the single-sample $\<\drec^{s'}\>_k$ in the denominator of the estimator \refeq{fourier_correlation}. This ensures that, if the inferred initial conditions are drawn from a posterior centered on the true initial conditions, $\rri(k)$ indeed asymptotes to 1. This would not be the case if one used $\<\drec^{s'}\>_k$ in the denominator, since the inference noise would then always lead to an $\rri(k)$ below 1.

We use the Gaussian expectation of $\rri$ as a reference. This large-scale (low-$k$) limit corresponds to the case wherein $\dm$, $\dh$, and $\varepsilon_h$ (halo stochasticity) fields are all \emph{Gaussian} fields, where the tracer and matter fields are linearly related.
As derived in \refapp{Gaussian_limit_of_rk}, the Gaussian expectation is given by:
\be
r_{\mathrm{ri},\lin}(k) = \frac{1}{\sqrt{1+P_\varepsilon(k)/\left(b_1^2\Plin(k)\right)}},
\label{eq:gauss_limit_rri}
\ee
where $\Plin(k)$ and $P_\varepsilon(k)$ denote the linear matter power spectrum and the halo stochasticity power spectrum, respectively.

\refeq{gauss_limit_rri} clearly shows that there is an expected limit for the inference, set by halo stochasticity and bias, if the inference only has access to information of the linear evolution of LSS.
Additionally, \refeq{gauss_limit_rri} also suggests that one expects $r_{\mathrm{ri},\lin}$ to asymptote to unity on large scales (low-$k$ values) and to decrease at progressively smaller scales (higher-$k$ values). This has nothing to do with non-linear evolution, but just the fact that $P_\varepsilon(k)$ asymptotes to a constant on large scales while $\Plin(k)$ is a decreasing function of $k$.
Note that at very low $k$ (below the range shown here), $r_{\mathrm{ri}, \lin}$ actually shrinks again due to the turnover in $\Plin(k)$.
For comparison purpose, since the halo stochasticity power spectrum cannot be precisely determined, we approximate \refeq{gauss_limit_rri} by the correlation coefficient between the actual halo and the input initial Fourier modes, $\rhi$, which is given by (cf. \refeq{fourier_correlation_general}):
\be
\rhi(k) = \frac{\<\dh \dinput\>_k}{\sqrt{\<\dh \dh\>_k\,\<\dinput \dinput\>_k}}.
\label{eq:rk_ht}
  \ee
Below, we will show \refeq{rk_ht} as dotted lines in all of our figures of correlation coefficients to serve as a reference. Note that \refeq{rk_ht} can deviate from \refeq{gauss_limit_rri} at high $k$, where the fields are significantly non-Gaussian. However, at low $k$, for whatever weighting applied (on the halos), \refeq{gauss_limit_rri} always reduces to \refeq{rk_ht}.

The transfer function can be defined for each MCMC sample $s$ as
\be
T_{\rec}^{s}(k) = \sqrt{\frac{P^{s}_{\rec}(k)}{P_{\true}(k)}},
\label{eq:transfer_function}
\ee
where we define the matter power spectrum as $\<\d(\vk)\d^\ast(\vk')\>_k\equiv(2\pi)^3\d_D\left(\vk-\vk'\right)P(k)$. The \borg ensemble mean and variance are then given by:
\be
T_{\rec}(k) = \<\sqrt{\frac{P_{\rec}(k)}{P_{\true}(k)}}\>_s,
\label{eq:Trec_mean}
\ee
\be
\sigma^2_T(k) = \frac{1}{N_{\mathrm{samples}}}\sum_{s}^{N_{\mathrm{samples}}}  \left[T_{\rec}^{s}(k)-T_{\rec}(k) \right]^2 .
\label{eq:Trec_var}
\ee
The mean inferred amplitudes of the initial conditions at different wavenumbers are unbiased compared to the truth if $T_{\rec}(k)$ is consistent with 1 within the uncertainty.

Below we vary each element in \refsec{setup} that could potentially affect the quality of the inference, while keeping the rest of the inference setup fixed.
We classify our results into two categories: those from inferences using mock data and those from inferences using N-body DM halo catalogs. For additional clarity, we explicitly specify the details of the setup at the beginning of each section.
 
\subsection{Results with mock data}
\label{sec:results_mock}

In addition to the N-body halo datasets described in \refsec{tracers}, we generate mock \enquote{halo} fields consistently following the same procedure of the forward inference:
\begin{enumerate}[leftmargin=1.5cm]%
\item Draw a Gaussian realization of $\d_{m,\ini}$ from \refeq{multivariate_Gaussian_ICs}.
\item Forward evolve $\d_{m,\ini}\to\d_{m,\fwd}$, assuming a certain gravity model (LPT or 2LPT).
\item Transform $\d_{m,\fwd}\to\d_{h,\det}$, assuming one of the bias models presented in \refsec{biasmodel}, plus some fiducial values for the parameters of that model.
\item Draw a random realization of $\P\left(\nh^i\Big|\nhdet^i\right)$, assuming either the Poisson or Gaussian likelihood presented in \refsec{likelihood}, plus some fiducial values for $n_0$ (and $\sigma$ in the case of the Gaussian).
 \end{enumerate}
 We use these mock datasets in \refsec{results_tracer_density} and \refsec{results_grid_resolution} below, each as input for an inference employing exactly the same choices of gravity model, bias model and likelihood. The goal is to isolate and study the effect of tracer density and grid resolution, respectively, on the inference.

\subsubsection{Tracer number density}
\label{sec:results_tracer_density}

All results in this section are obtained with the following setup:
\begin{itemize}
	\item 2LPT forward model.
	\item Power-law bias model.
	\item Poisson likelihood.
	\item $\lgrid=15.6\,\Mpch$.	
\end{itemize}
We consider three mock \enquote{halo} catalogs, as described at the beginning of this section.
We emphasize again that, as the biased field is generated from the exact models used in the inference, any difference between the qualities of different inferences should really arise from the tracer number density alone (the same applies for the grid resolution examined in \refsec{results_grid_resolution} below).
We consider three cases with tracer (comoving) number densities of $\avnh=3.34\times10^{-4},\, 1.10\times10^{-4},\,2.91\times10^{-5}\,(\iMpch)^3$. In \reffig{rk_ek_tracer_density}, we show results for these inferences.

\begin{figure*}[thbp]
	\centering
	\samelinetwofig{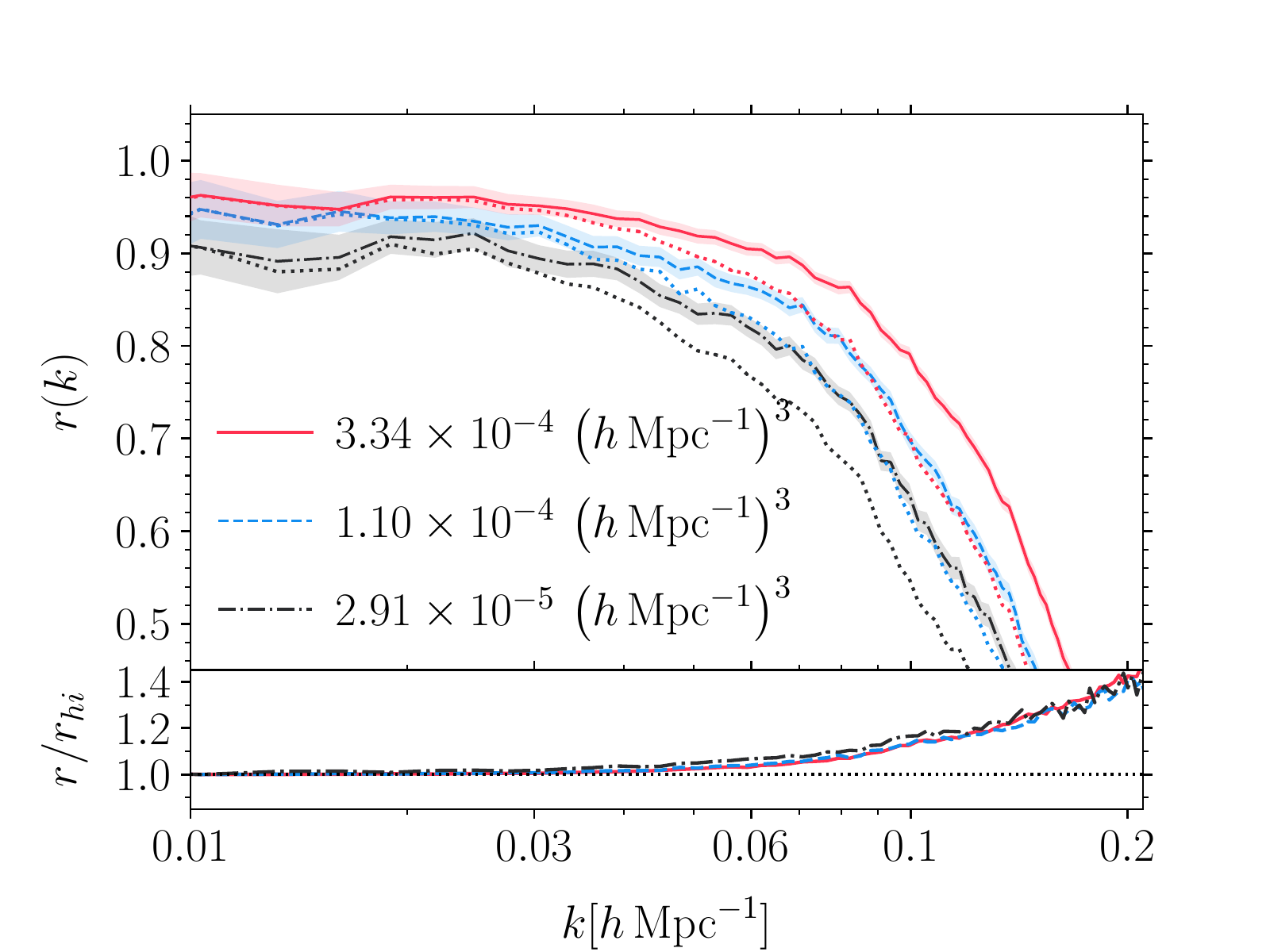}{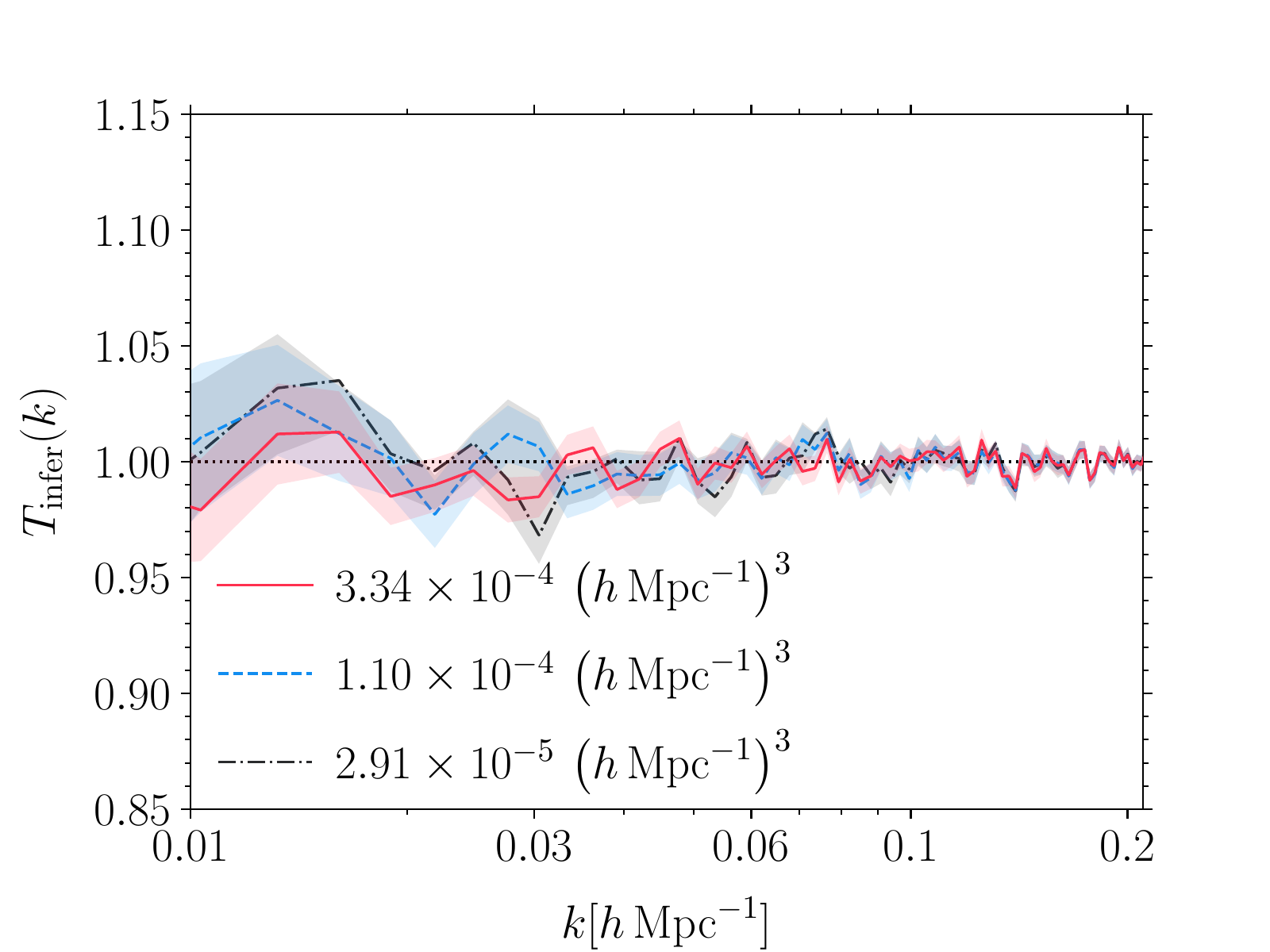}{fig:rk_ek_tracer_density}{\textbf{Varying tracer density - mock data.} \textit{Left:} The phase-correlation coefficient, estimated by \refeq{rri_mean}, between the inferred and the true initial conditions, for different tracer densities, plotted in solid lines. The corresponding shaded bands denote the 1-$\sigma$ uncertainties. For reference, the large-scale, Gaussian expectations (cf. \refeqs{gauss_limit_rri}{rk_ht}) are plotted in dotted lines. \textit{Right:} The amplitude-bias between inferred and true initial conditions, measured in terms of the transfer function $T_{\rec}$, for the same cases shown on the left. The solid lines again represent the ensemble means (cf. \refeq{Trec_mean}) while the corresponding shaded bands designate the 1-$\sigma$ uncertainty (cf. \refeq{Trec_var}).}
\end{figure*}

As shown in the left panel, upper plot of \reffig{rk_ek_tracer_density}, the phase-correlation coefficient is \emph{scale-dependent}. At moderately high $k$, the inference from mock data performs slightly better than the corresponding Gaussian expectation, while it perfectly agrees with the latter at low $k\leq0.04\,\iMpch$. This trend indicates two things:
\begin{itemize}
	\item The forward inference does have access to information encoded in the non-linear evolution of LSS and can recover the small-scale phases that have been processed by gravity in the mildly non-linear regime. The inference quality relative to the Gaussian expectation depends very weakly on the tracer density, at least on the scales considered here.
	\item The Gaussian expectation correctly describes the phase inference on large scales, even in the absence of modeling error.
\end{itemize}

Indeed, the bottom plot in the left panel of \reffig{rk_ek_tracer_density} , which shows the ratios between the phase-correlation coefficients and their corresponding Gaussian expectations, confirms that the Gaussian expectation is an excellent indicator of the cross-correlation between true and inferred initial conditions at low $k$ for a wide range of tracer densities.

\begin{figure*}[thbp]
     \centering
     \twolinefourfig{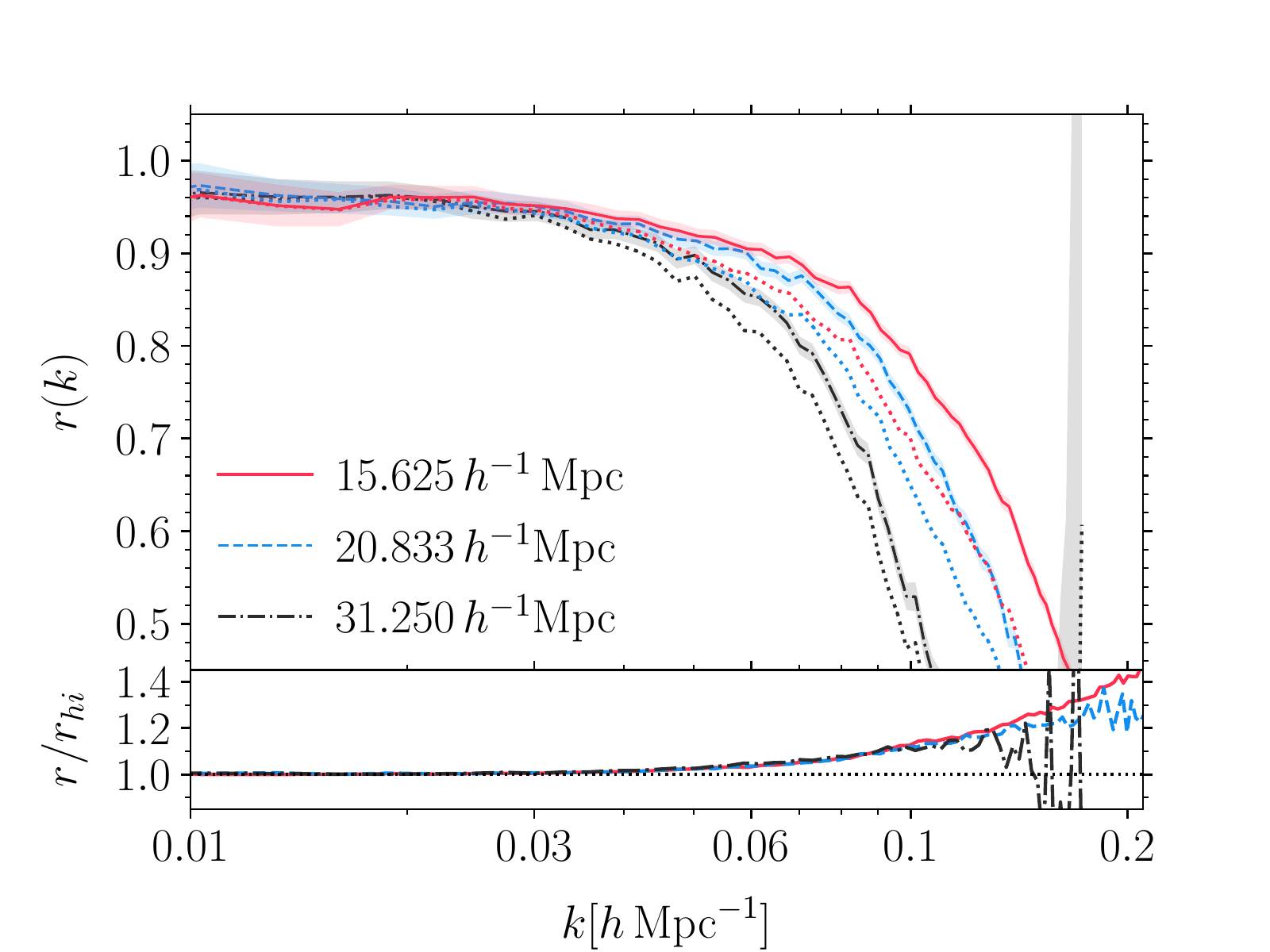}{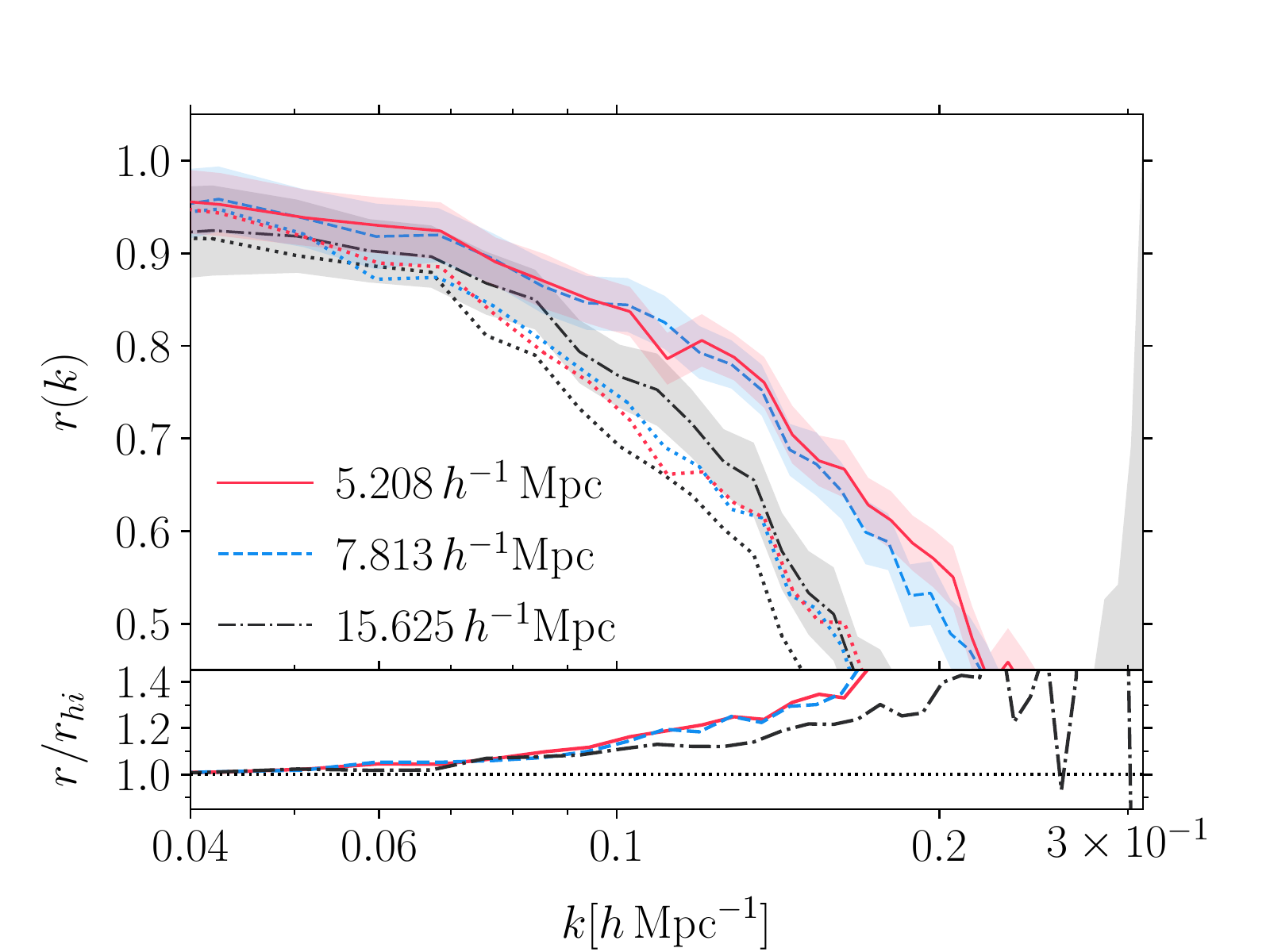}
     {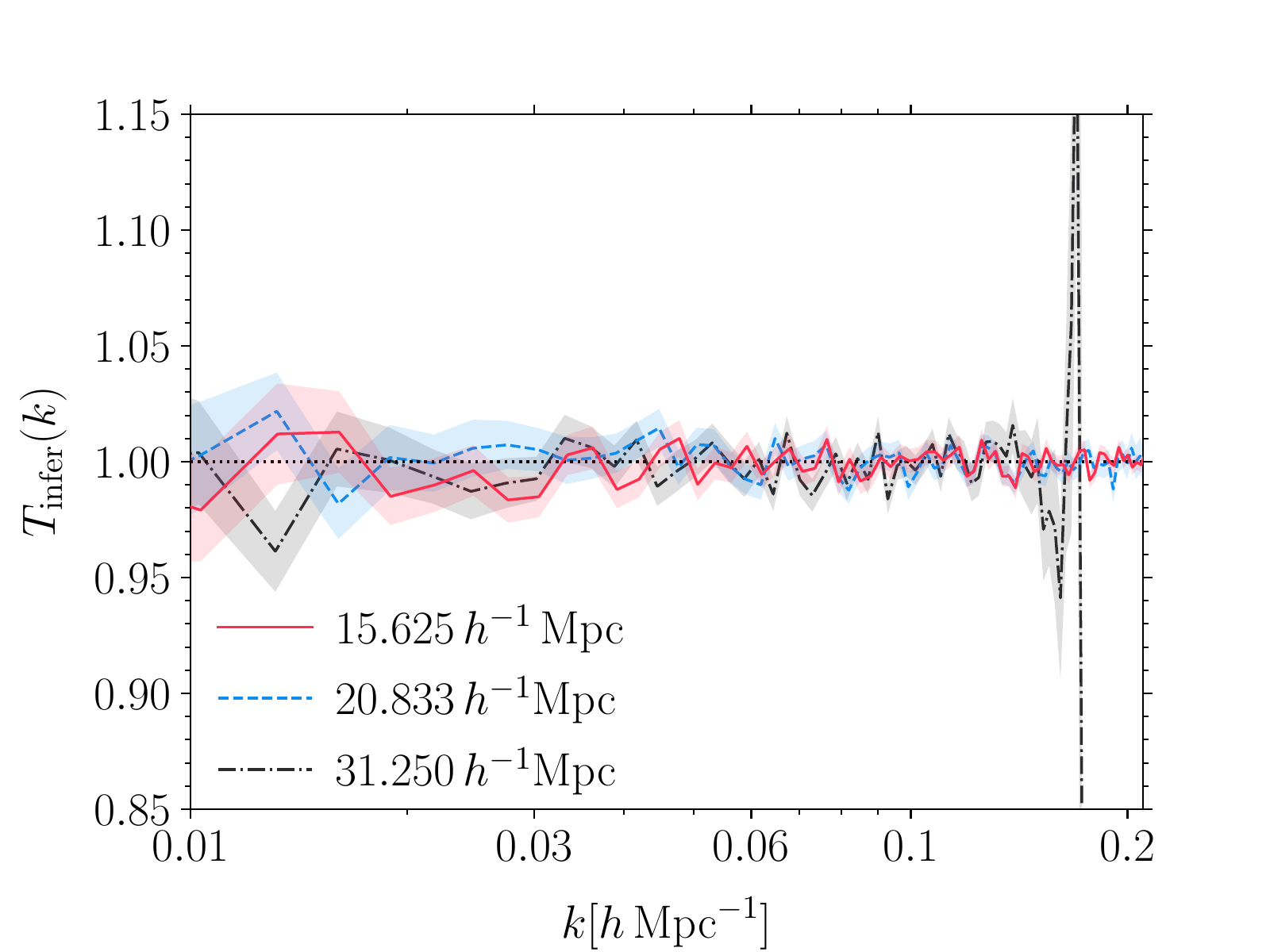}
     {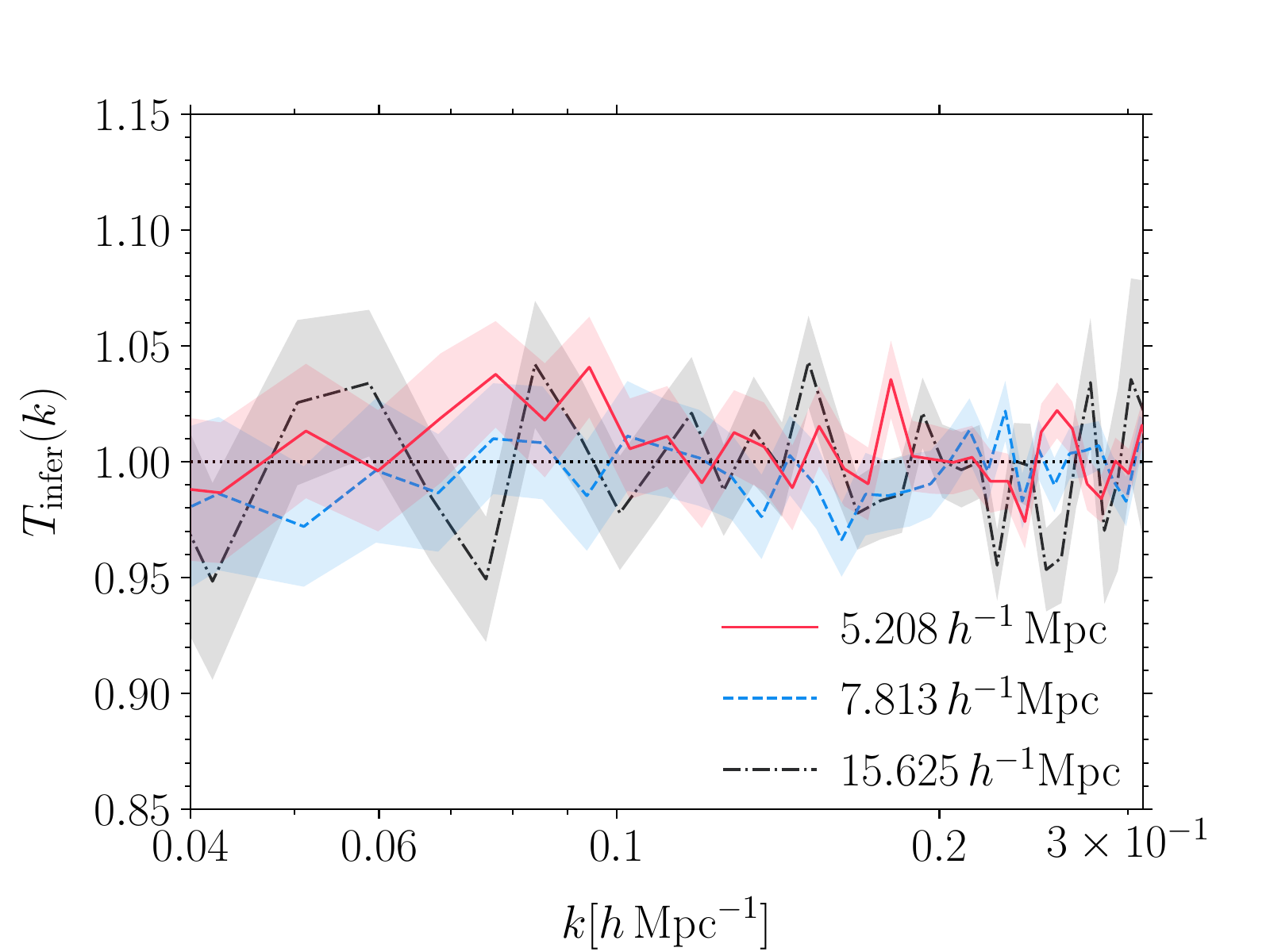}{fig:rk_ek_grid_resolution}{\textbf{Varying grid resolution - mock data.} \textit{Top:} The phase-correlation coefficient $r(k)$ in inferences with $\lgrid=31.3,\,20.8,\,15.6$ $\Mpch$ (left) and $\lgrid=15.6,\,7.8,\,5.2$ $\Mpch$ (right). Dotted lines again represent the Gaussian expectations. \textit{Bottom:} The transfer function $T_{\rec}(k)$ in the corresponding inferences.}
\end{figure*}

The right panel of \reffig{rk_ek_tracer_density} shows the amplitude-bias, or transfer function, in the corresponding inferences. The transfer functions are consistent with 1, which demonstrates the consistency of the inference pipeline. The right panel of \reffig{rk_ek_tracer_density} shows a direct comparison between transfer functions in three inferences, which interestingly appear to be different only at $\lesssim2$\% level at the lowest $k$ values.
Another important observation is that the uncertainties, \emph{introduced by marginalizing over the Fourier phases}, is of order of $\lesssim2$\% for both the phase-correlation and amplitude-bias. This shows that there is still a lot of information and constraining power even when the phases are sampled, as done in the field-level, forward-modeling approach.

We emphasize again, however, that there is no modeling error since the mock datasets employed in these inferences are sampled from the model assumed in the inference.
Additionally, the three mock datasets were generated with the same phases, which explains the correlated fluctuations in the three different mock tracer cases at high $k$ in the right panel of \reffig{rk_ek_tracer_density}.

\subsubsection{Grid resolution}
\label{sec:results_grid_resolution}

The general setup in this section is as follows:
\begin{itemize}
	\item 2LPT forward model.
	\item Power-law bias model.
	\item Poisson likelihood.
	\item Mock \enquote{halos} with $\avnh=3.34\times10^{-4}\,(\iMpch)^3$ in one $\Lbox=2000\,\Mpch$ (mimicking \boxone) and one $\Lbox=500\,\Mpch$ box (mimicking \boxtwo).
\end{itemize}
We run three inferences on each mock data set, at five different grid resolution settings: $\lgrid=31.3,\,20.8,\,15.6$ $\Mpch$ for the bigger box, and $\lgrid=15.6,\,7.8,\,5.2$ $\Mpch$ for the smaller one.
Increasing the grid resolution certainly increases the amount of input information. However, as discussed in \refsec{notation}, one must be careful not to include information from highly non-linear scales where our model is no longer valid. Fortunately, with mock data, we can just focus on the effect of the former, i.e. being able to access more small-scale information, on the inference.

The top panels of \reffig{rk_ek_grid_resolution} compare the phase-correlation coefficients from six inferences with five different grid resolutions. Higher resolutions are observed to considerably improve $r(k)$ at high-$k$ values.
Interestingly, the improvement when going from $\lgrid=7.8\,\Mpch$ to $\lgrid=5.2\,\Mpch$ is marginal. This is presumably because we are reaching the shot noise dominated regime, where most cells are empty for this tracer density.
We show, in the bottom panels of \reffig{rk_ek_grid_resolution}, a comparison between transfer functions from the corresponding inferences. Similarly to the case of varying tracer density, the transfer function appears to be rather insensitive to grid resolution. Thus, for such cases where the physical data model perfectly matches the underlying mechanism generating the data, the inference is robustly unbiased.

\subsection{Results with N-body data}
\label{sec:result_GADGET2}

Results shown in \reffig{rk_ek_tracer_density} and \reffig{rk_ek_grid_resolution} are obtained with mock data, for which there is no mismatch between the mechanisms generating the data and the models employed in the inference. When N-body halos from \boxone are instead taken as input data in each of the setup shown above, i.e. all inference settings being the same, we obtain the results shown in \reffig{rk_ek_tracer_density_grid_resolution_halo}.
These results essentially confirm the trends observed with mock data with one interesting exception, to which we will turn shortly.

First, for the phase-correlation, we observe the same trends, namely the scale-dependence behavior and the sharp decrease after some characteristic wavenumber. Despite possible differences between the physical data model and the actual data, the inferences continue to outperform the Gaussian expectations. 
Second, for the transfer function, we find no strong dependence on either tracer density or grid resolution. Interestingly, we observe---consistently across all inferences---a clear \emph{scale-dependent} bias which cannot be explained by the scale-dependent, percent-level bias introduced by different normalization conventions between CLASS (CAMB) and \borg (see \refapp{normalization_bias} for more details).

The origin of this bias in the measured $T_{\rec}(k)$ is perhaps easier to understand in the linear regime, where
\ba
\d_{m,\fwd}(\vk) &= D_+(a) \d_{m,\ini}(\vk),\\
\dh(\vk) &= b_1 \d_{m,\fwd}(\vk) = D_+(a)\, b_1 \d_{m,\ini}(\vk),
\ea
with $D_+(a)$ being the linear growth factor.
One immediately sees that the linear bias $b_1$ and the amplitude of initial fluctuations are perfectly degenerate at linear level. A wrong inferred $b_1$ (or $\beta$ in the context of the power-law bias model) thus entails a bias in the amplitude of the inferred modes. To correctly recover both bias parameter(s) and the amplitude, the physical data model must be flexible enough while still keeping sub-grid systematics under rigorous control \cite{Schmidt:2020b}.
In particular, the fact that the bias in $\beta$ only shows up with N-body input further implies that this bias is of physical origin, and thus can only be addressed by improving either the gravity model, the bias model, the likelihood, or any combination of these three ingredients of the physical data model.
Below, we repeatedly return to this point in \refsec{results_forward_model}-\refsec{results_likelihood}. We further show that it is highly likely \emph{a combination of the bias model and likelihood} (and to some extent, the gravity model) that is in charge and therefore should be improved.

\begin{figure*}[thbp]
	\centering
	\twolinefourfig{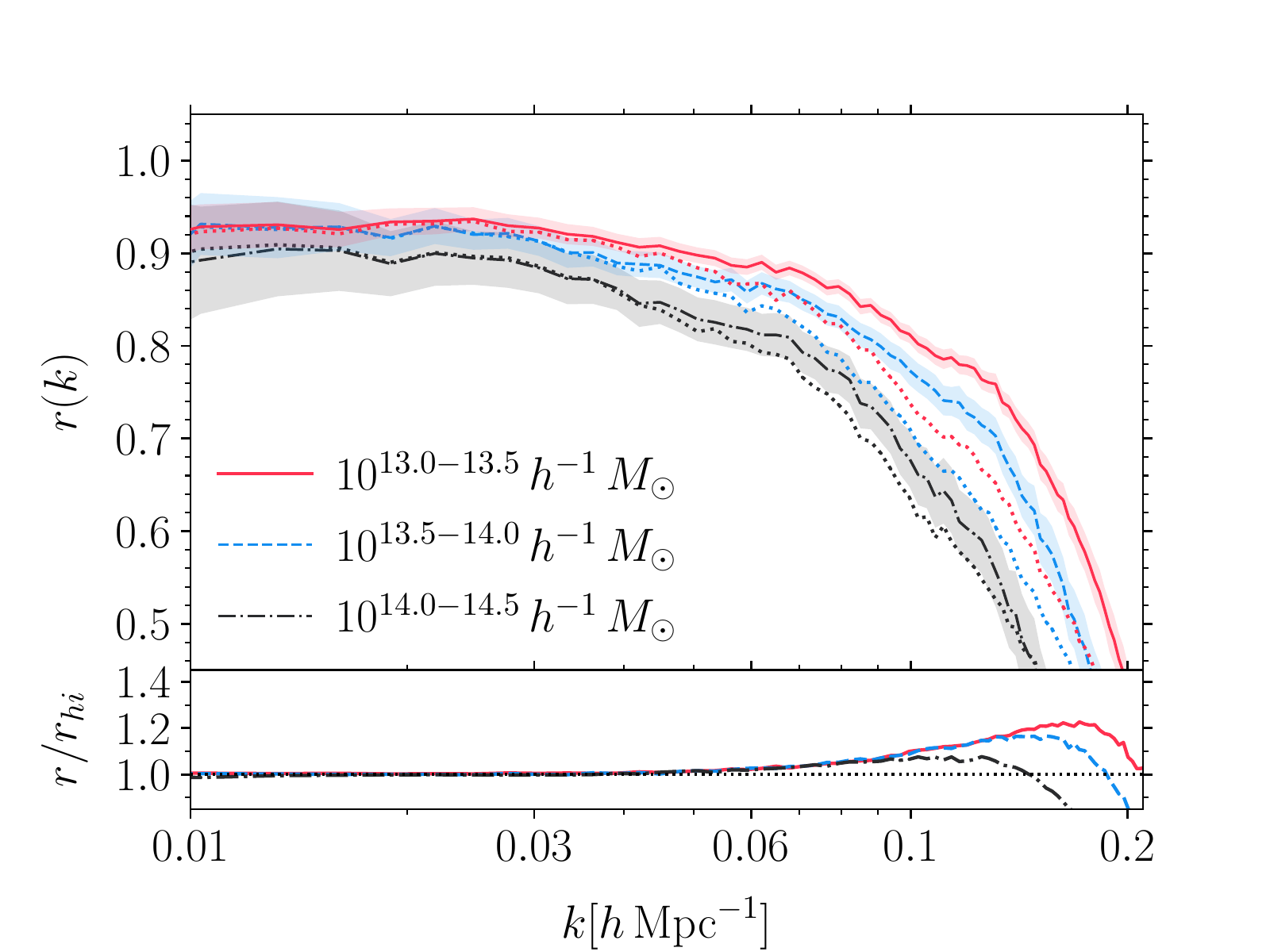}{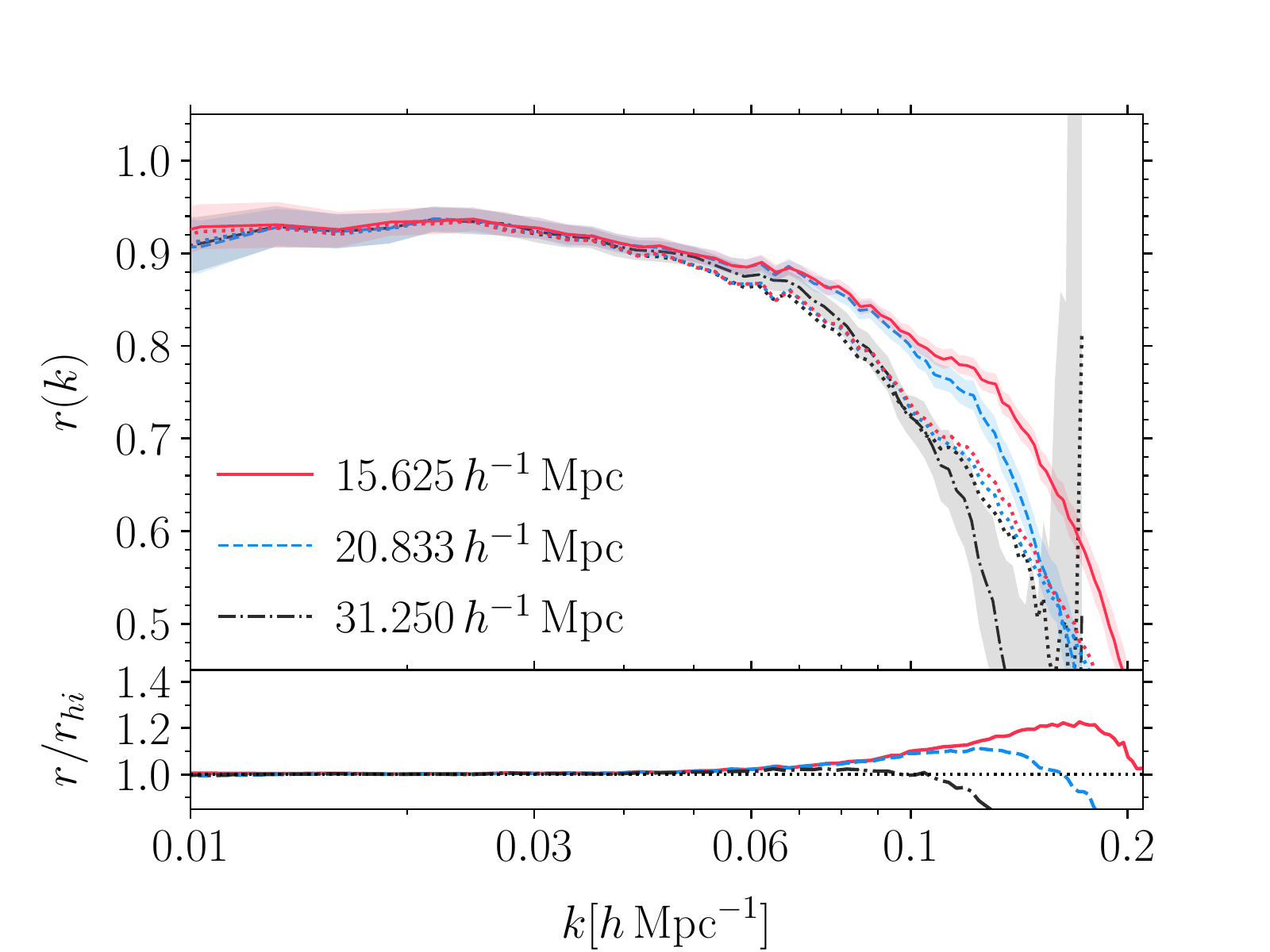}
	{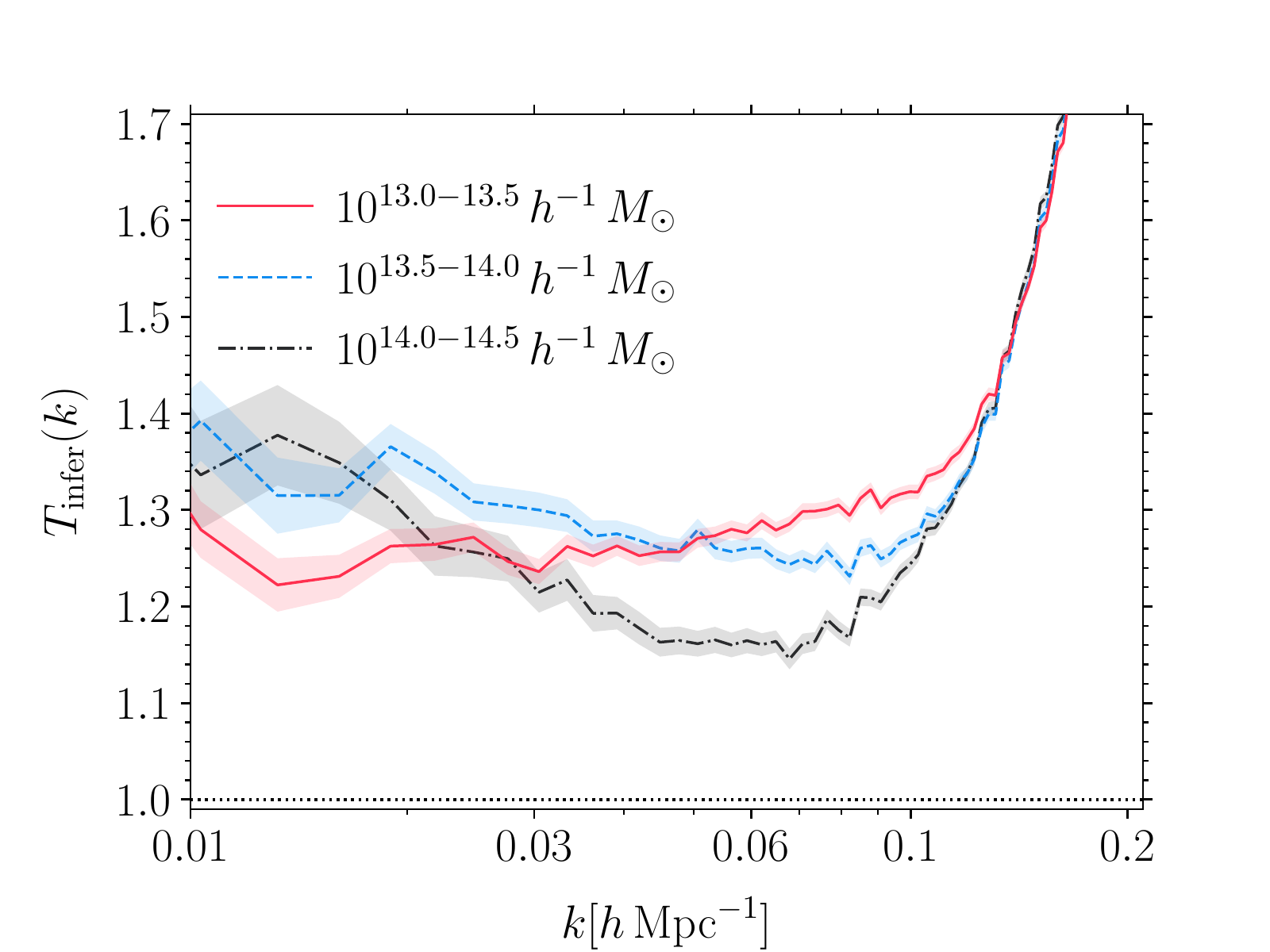}
	{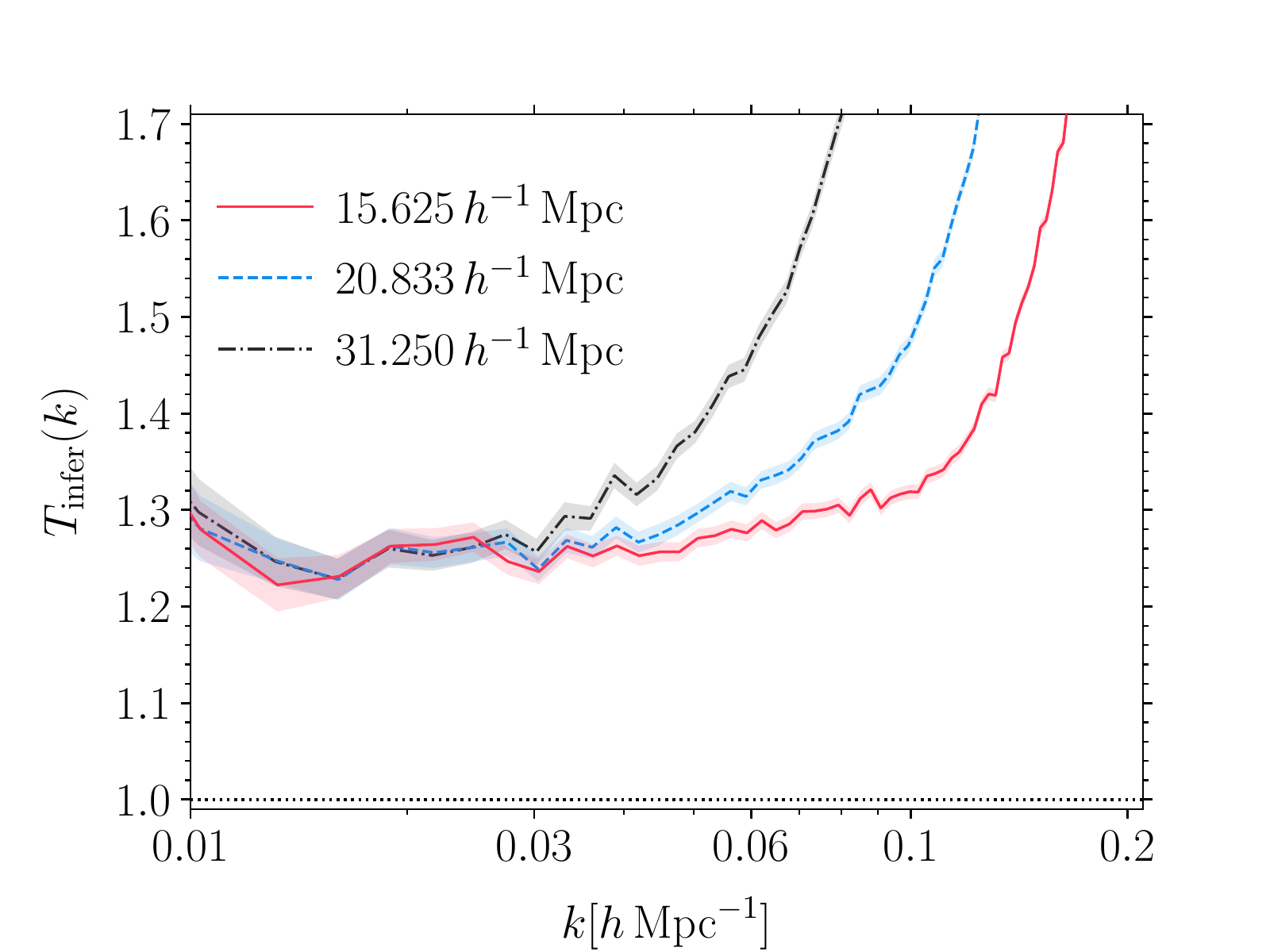}{fig:rk_ek_tracer_density_grid_resolution_halo}{\textbf{Varying tracer density \& grid resolution - N-body data.} \textit{Top:} The phase-correlation coefficient in different inferences with varying tracer density (left) and grid resolution (right), when \boxone is taken as input data of the inference. The Gaussian expectations, plotted in dotted curves, serve as a reference for the performance of each inference. \textit{Bottom:} The transfer function in the corresponding inferences.}
\end{figure*}

\subsubsection{Gravity model}
\label{sec:results_forward_model}

To obtain results in this section, we use the following setup:
\begin{itemize}
	\item Power-law bias model.
	\item Poisson likelihood.
	\item \boxone (fiducial sample) and \boxtwo.
	\item $\lgrid=15.6\,\Mpch$ (\boxone) and $\lgrid=5.2\,\Mpch$ (\boxtwo).
\end{itemize}
We run two inferences using LPT and 2LPT as the forward model. We note that LPT speeds up the HMC sampler by approximately a factor of 3, on average. The warm-up phases, during which the chains approach the stationary target distribution (see \refapp{burnin_and_thinning}), are very similar in both cases in terms of number of MCMC samples required.

As shown in \reffig{rk_ek_forward_model}, the performances of inferences using LPT mirror those of inferences using 2LPT extremely well at $\lgrid=15.6\,\Mpch$, in terms of both metrics.
The ensemble means of the inferred initial density fields using LPT and 2LPT are $\gtrsim99\%$ spatially-correlated. We further verify that this trend is universal among inferences using different bias models and likelihoods.
To test the limit of LPT, we increase the resolution to $\lgrid=5.2\,\Mpch$ using \boxtwo. In this case we observe a 5-10\% difference in performance, favoring 2LPT.
Taking into consideration also the computational demand, these results jointly suggest that, for large-volume surveys where the grid resolution will be anyway limited (due to numerical restrictions on number of grid cells), LPT could be a sufficient choice. In fact, the validity of LPT as the forward model for an inference has already been demonstrated in the BOSS-SDSS3 inference in \cite{Lavaux:2019} (see Section 4.4, especially figure 6).
The limit $\lgrid=5.2\,\Mpch$---where LPT starts to fall behind 2LPT---however, should be noted.

Going from LPT to 2LPT shifts the transfer functions towards unity, which is the right direction. The shifts are however not nearly enough to yield transfer functions consistent with 1. We therefore argue that the gravity model is not the key to this issue. 

\begin{figure*}[thbp]
	\centering
	\twolinefourfig{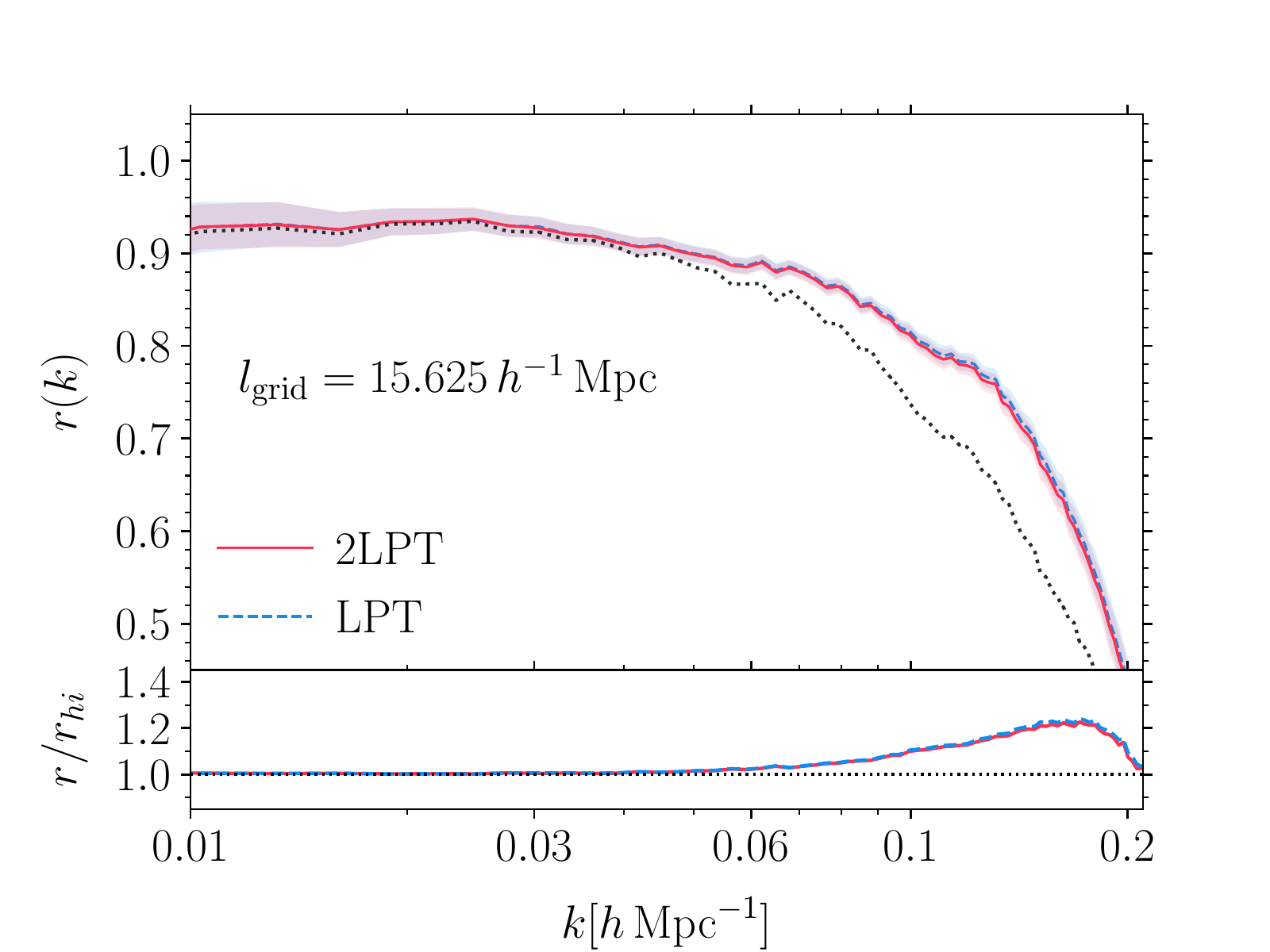}{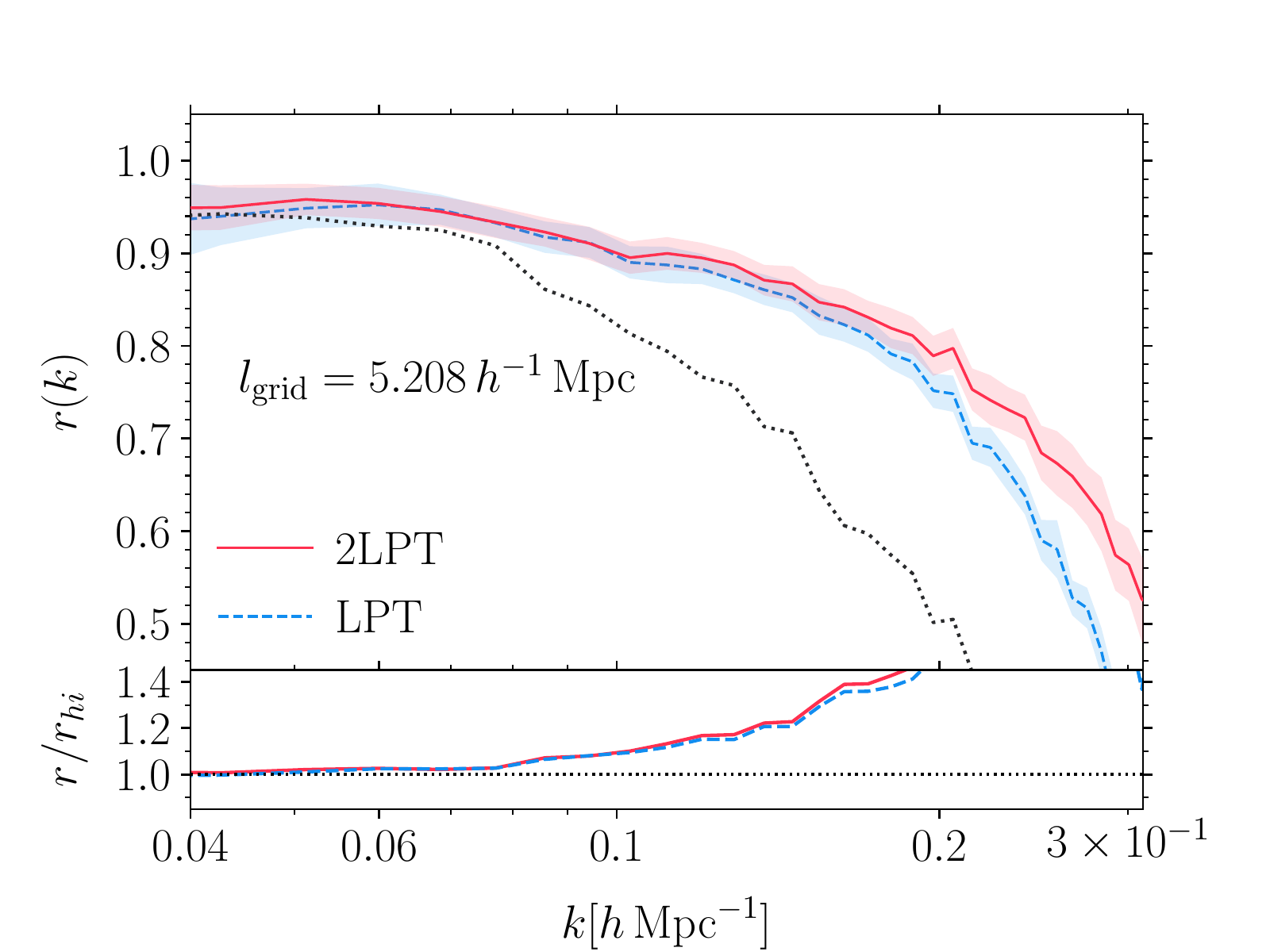}
	{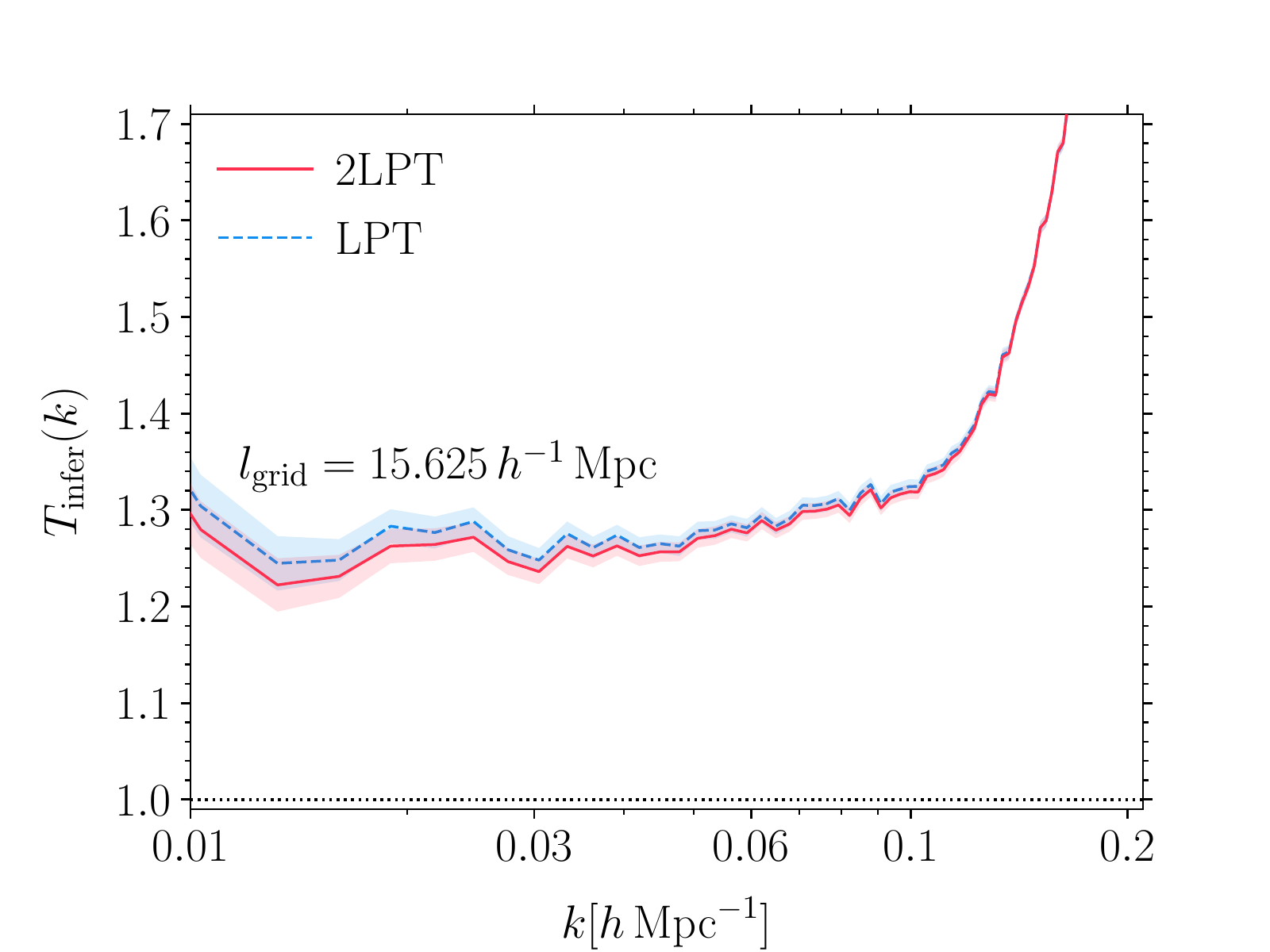}
	{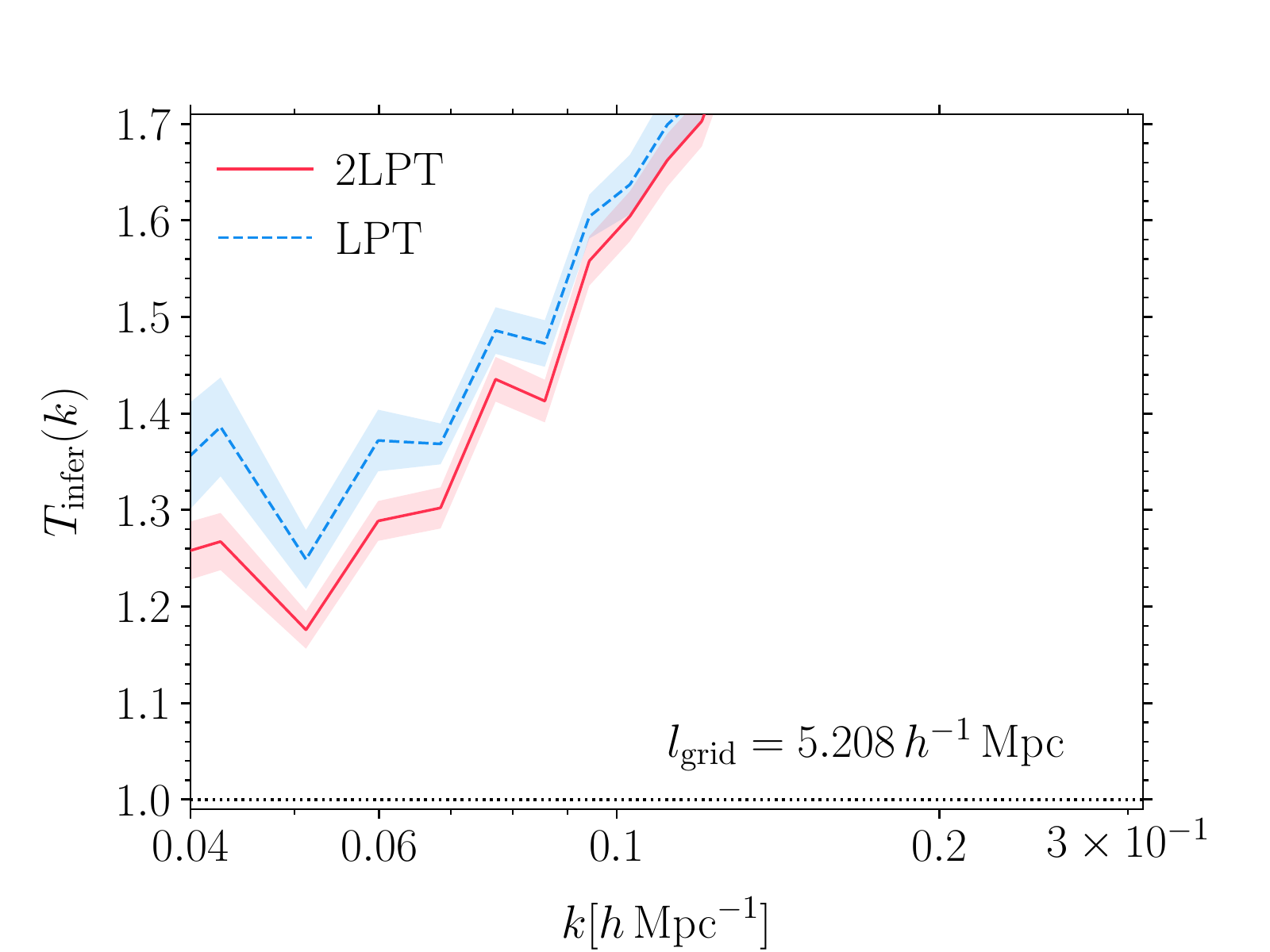}{fig:rk_ek_forward_model}{\textbf{Varying forward model - N-body data.} \textit{Top:} The phase-correlation coefficient in different inferences with LPT and 2LPT at the grid resolution of $\lgrid=15.6\,\Mpch$ using \boxone (left) and $\lgrid=5.2\,\Mpch$ using \boxtwo (right). The dark-gray dotted line shows the Gaussian expectation, which is the same for both cases. \textit{Bottom:} The transfer function in the corresponding inferences.}
\end{figure*}

\subsubsection{Bias model}
\label{sec:results_bias_model}

In this section, we adopt the following general setup:
\begin{itemize}
	\item 2LPT forward model.
	\item Poisson likelihood.
	\item \boxone (fiducial sample) and \boxtwo.
	\item $\lgrid=15.6\,\Mpch$ (\boxone) and $\lgrid=5.2\,\Mpch$ (\boxtwo).
\end{itemize}
We consider three deterministic bias relations: the linear, the power-law, and the broken power-law models (cf. \refsec{biasmodel}).
For clarity, when combining linear bias model (cf. \refeq{linear_biasmodel}) and Poisson likelihood (cf. \refeq{cell_Poisson_likelihood}), we need to implement a soft-thresholder which truncates $\dhdet$ to ensure that $\nhdet\geq0$. Details about the thresholding procedure and priors on our bias parameters can be found in \refapp{prior_and_thresholder}.

\begin{figure*}[thbp]
	\centering
	\twolinefourfig{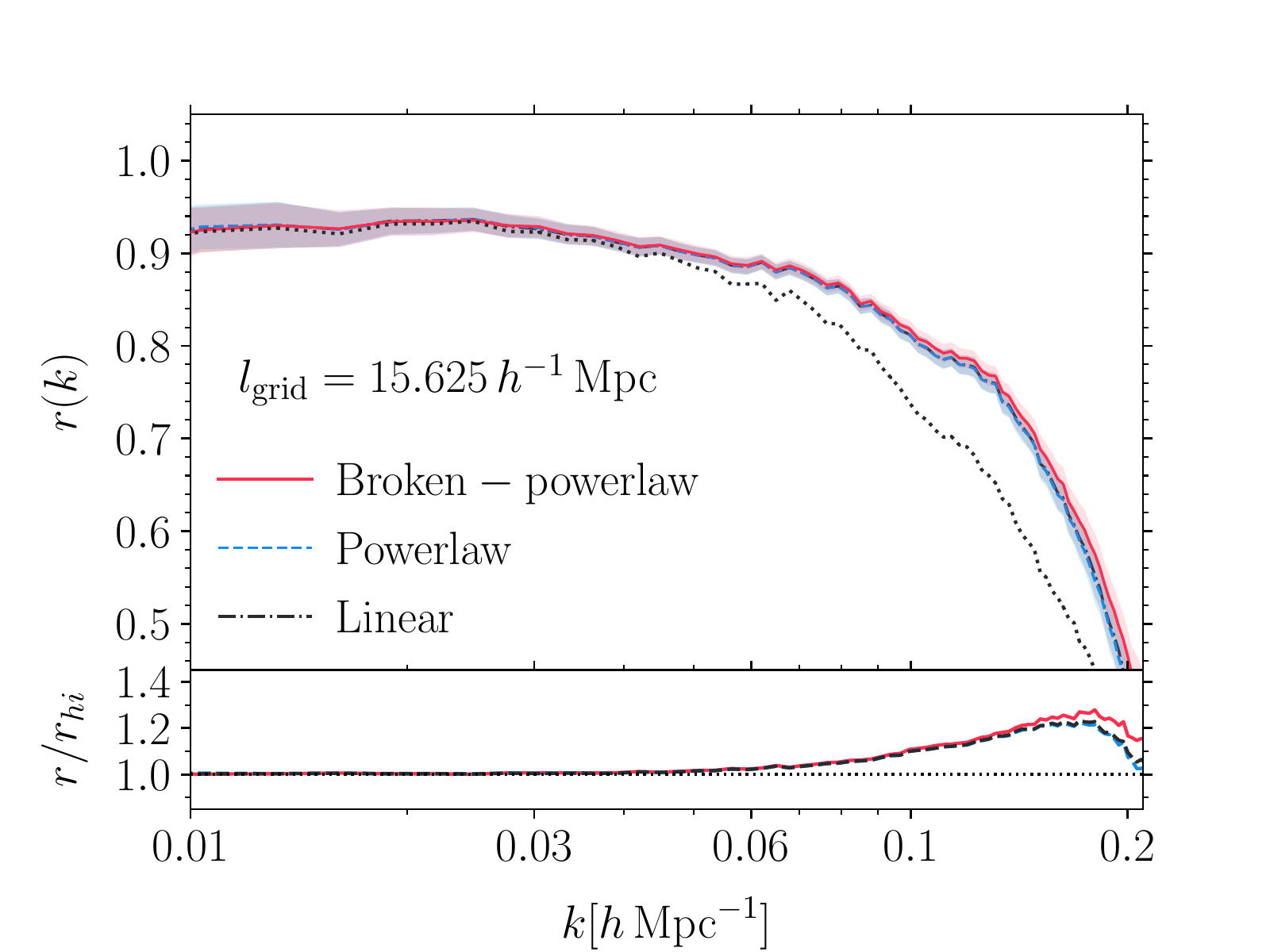}{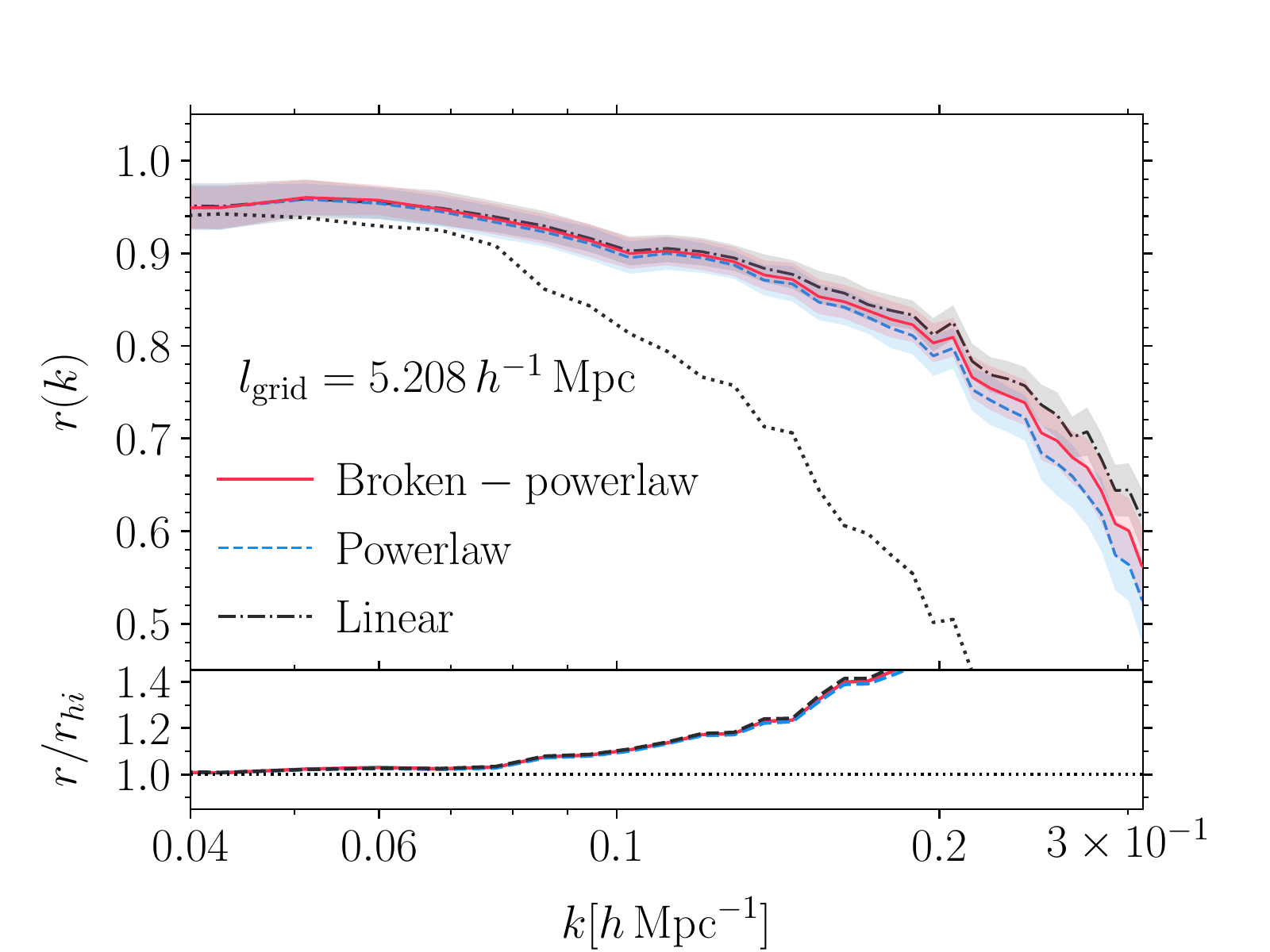}{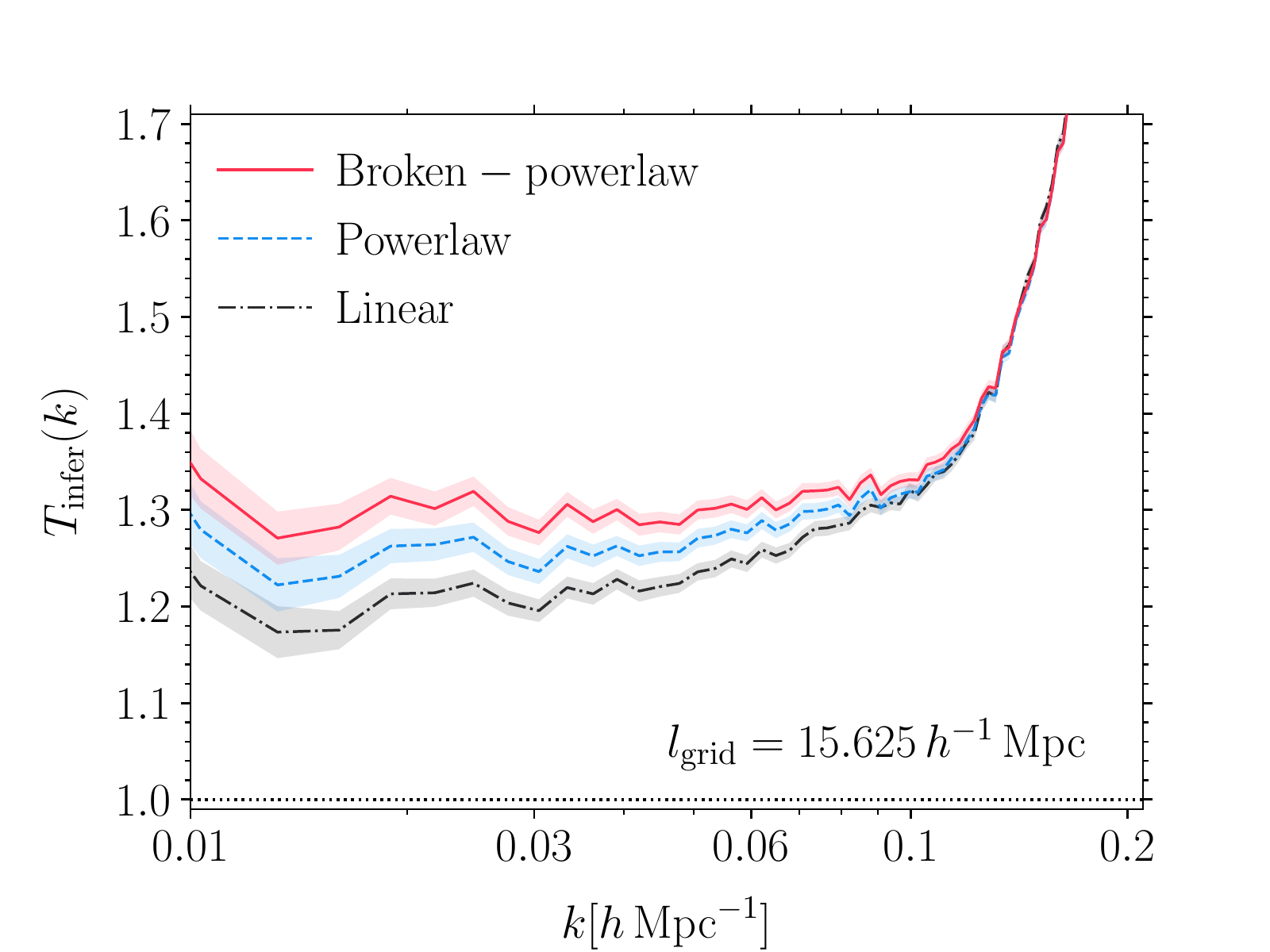}{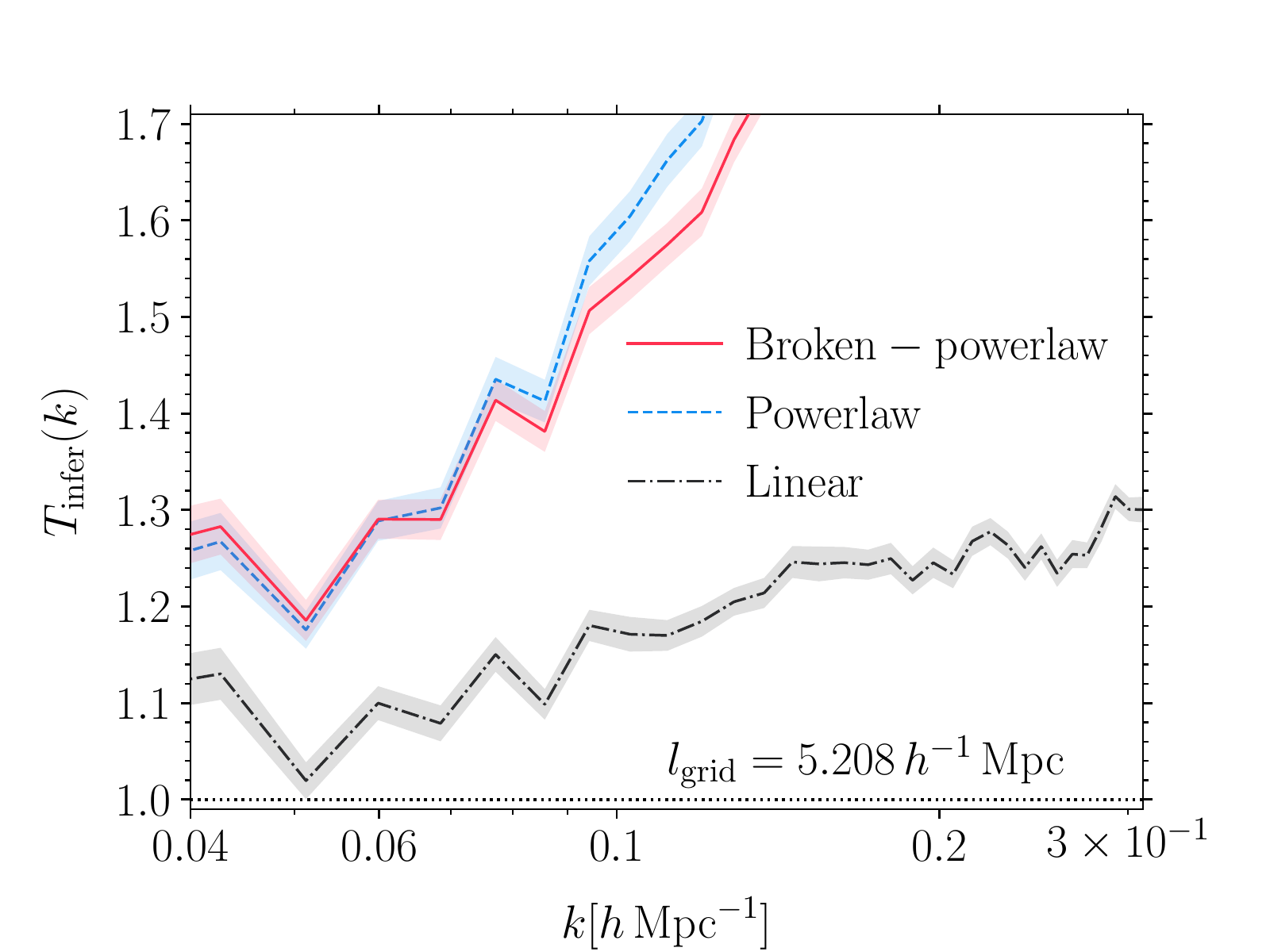}{fig:rk_ek_bias_model}{\textbf{Varying bias model - N-body data.} \textit{Top:} The phase-correlation coefficient in different inferences with different bias models described in \refsec{biasmodel}. The same Gaussian expectation for all three cases is illustrated by the dark-gray dotted line. \textit{Bottom:} The transfer function in the same inferences.}
\end{figure*}

\reffig{rk_ek_bias_model} shows results of inferences using the three bias models described above, for \boxone at $\lgrid=15.6\,\Mpch$ (top panels) and \boxtwo at $\lgrid=5.2\,\Mpch$ (bottom panels).
For the former, all bias models considered perform comparably in both metrics: The differences in $r(k)$ are all within the 1-$\sigma$ uncertainties; the differences in transfer functions favor the linear bias which exhibits a bias of only roughly 6-10\% less than the others.
More interestingly, results for the higher-resolution case clearly support this trend: both metrics favor the linear bias model. The differences in the transfer function now increase significantly. Still, none of the cases show a transfer function consistent with unity.

The fact that the differences between bias models are amplified with a higher grid resolution strongly indicates that it is the cells whose fluctuations have become non-linear that dominate the r.h.s of \refeq{powerlaw_biasmodel_approx}, and the expansion itself fails to converge at these resolutions.
Note that the linear bias model, as implemented with the Poisson likelihood here, is not truly linear. That is, the soft-thresholder, described in \refapp{prior_and_thresholder}, which ensures that $\nhdet^i$ is always positive (as required for a Poisson variable), naturally deals with voids and extremely low density regions by setting $\nhdet^i=0$ for those cells. A fairer comparison between the linear and power-law bias models requires, for example, a Gaussian likelihood. We return to this subtlety in \reffig{rk_ek_likelihood_linearbias} where it will become clear that the thresholding procedure applied in this section is highly unlikely to be the main reason for the superior performance of the linear bias model.

\subsubsection{Likelihood}
\label{sec:results_likelihood}

In this section, we adopt the following general setup:
\begin{itemize}
	\item 2LPT forward model.
	\item Power-law bias model.
	\item \boxone (fiducial sample) and \boxtwo
	\item $\lgrid=15.6\,\Mpch$ (\boxone) and $\lgrid=5.2\,\Mpch$ (\boxtwo).
\end{itemize}
We consider two forms of likelihoods: Poisson and Gaussian distributions (cf. \refsec{likelihood}).
In the case of the Gaussian likelihood, we find that letting the Gaussian variance $\sigma^2$ (cf. \refeq{cell_Gaussian_likelihood}) free is generally problematic for cases where the average Poisson variance of tracer counts-in-cell is small, that is
\be
\sigma_{\mathrm{Poisson}}^2\equiv\<\left(\nh^i-\nhdet^i\right)^2\>_{\mathrm{Poisson}}=N_h/\ngrid^3\ll1.
\label{eq:small_Poisson_variance}
\ee
Typically for such cases, we find that the variance parameter $\sigma^2$ collapses to a very small value, i.e. $\sigma^2\ll N_h/\ngrid^3$, and the chain only explores a narrow region around this value in posterior space. 
In order avoid this issue, we fix $\sigma^2=N_h/\ngrid^3$ in all inferences employing a Gaussian likelihood. 
It would be worth trying to alleviate this issue in the future by, for example, relaxing the universal variance assumption in \refeq{cell_Gaussian_likelihood} and allowing for the variance to be density-dependent, as discussed in \cite{Schmidt:2020b}\footnote{This is the case for the Poisson likelihood which specifically has a variance of $[\sigma^2]^i=\nhdet^i\equiv n_0(1+\dhdet^i)$}. This way, the likelihood could essentially downweight regions with low signal-to-noise while upweighting those with higher signal-to-noise.\footnote{The 2-point function analogy of which is the marked power spectrum \citep[e.g.][]{Massara:2020, Philcox:2020a}.}

\begin{figure*}[thbp]
	\centering
	\twolinefourfig{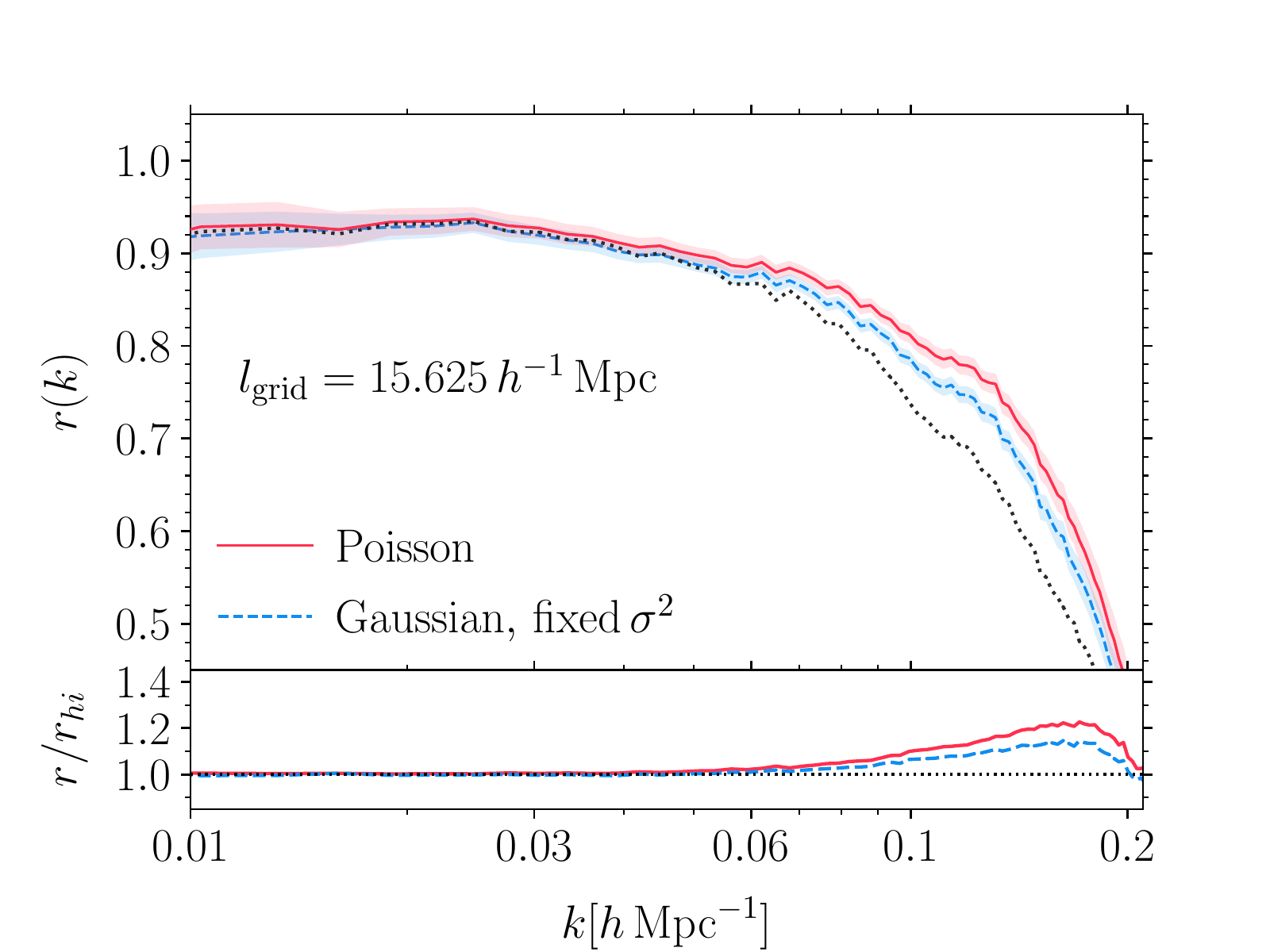}{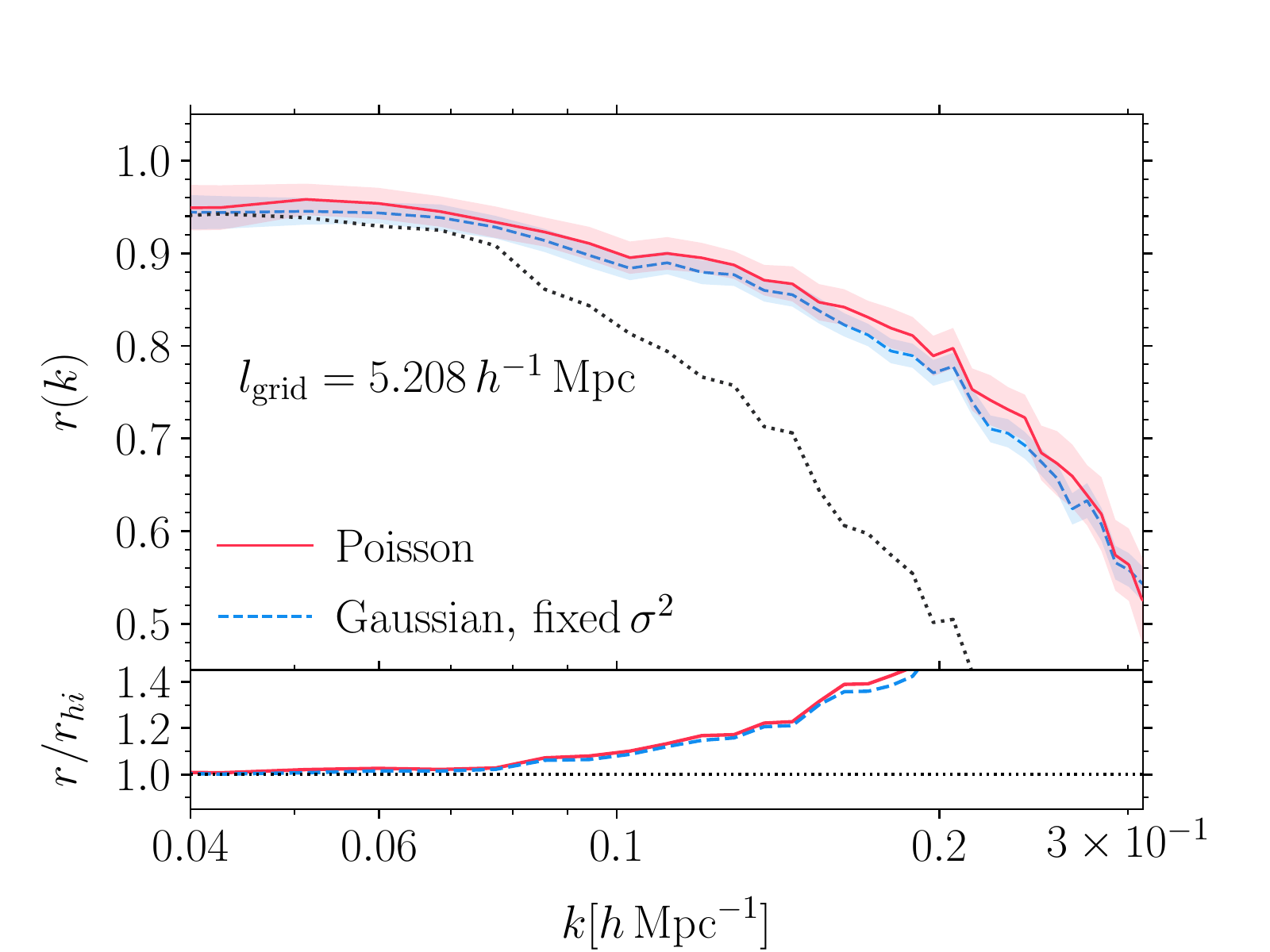}{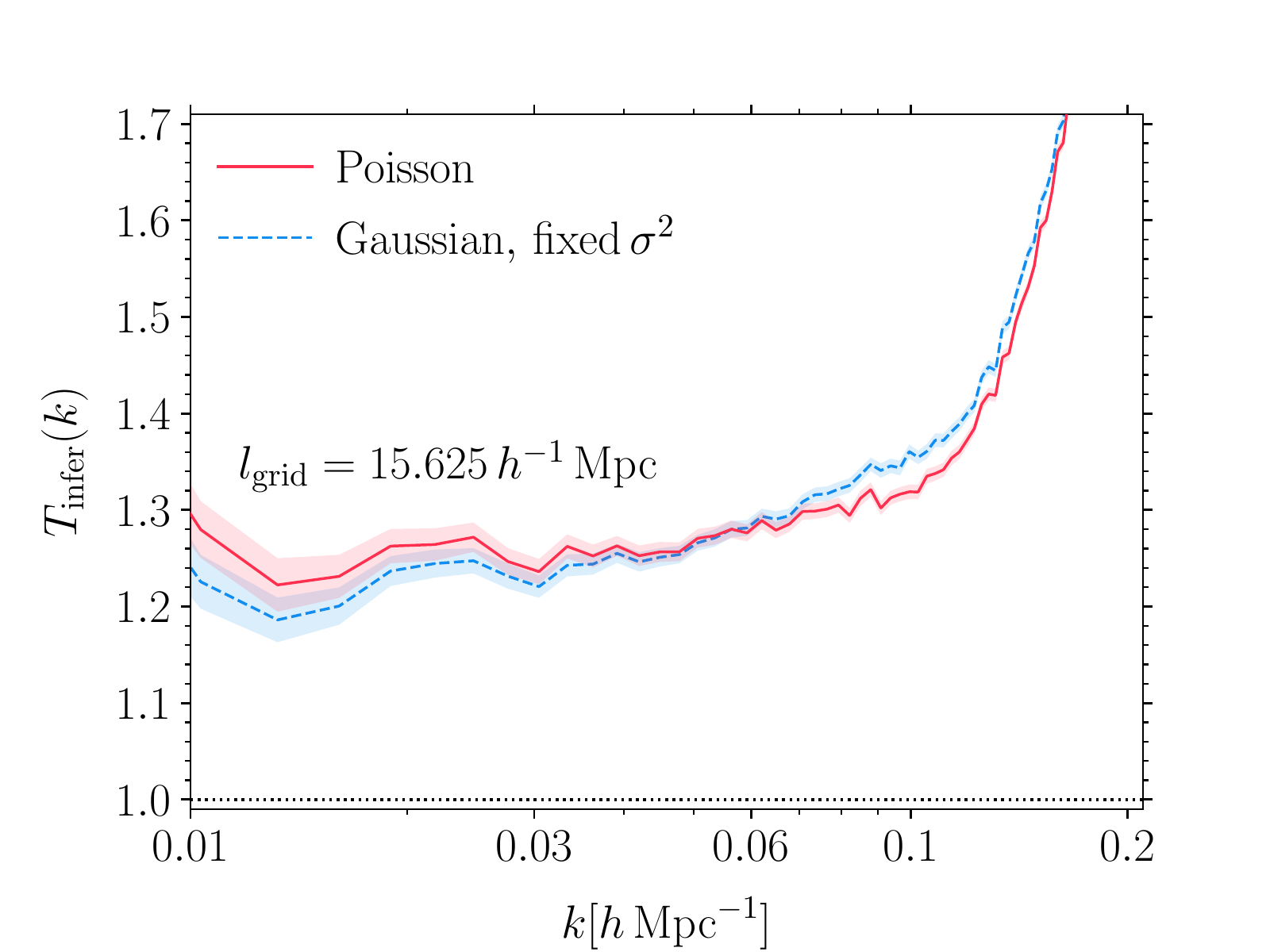}{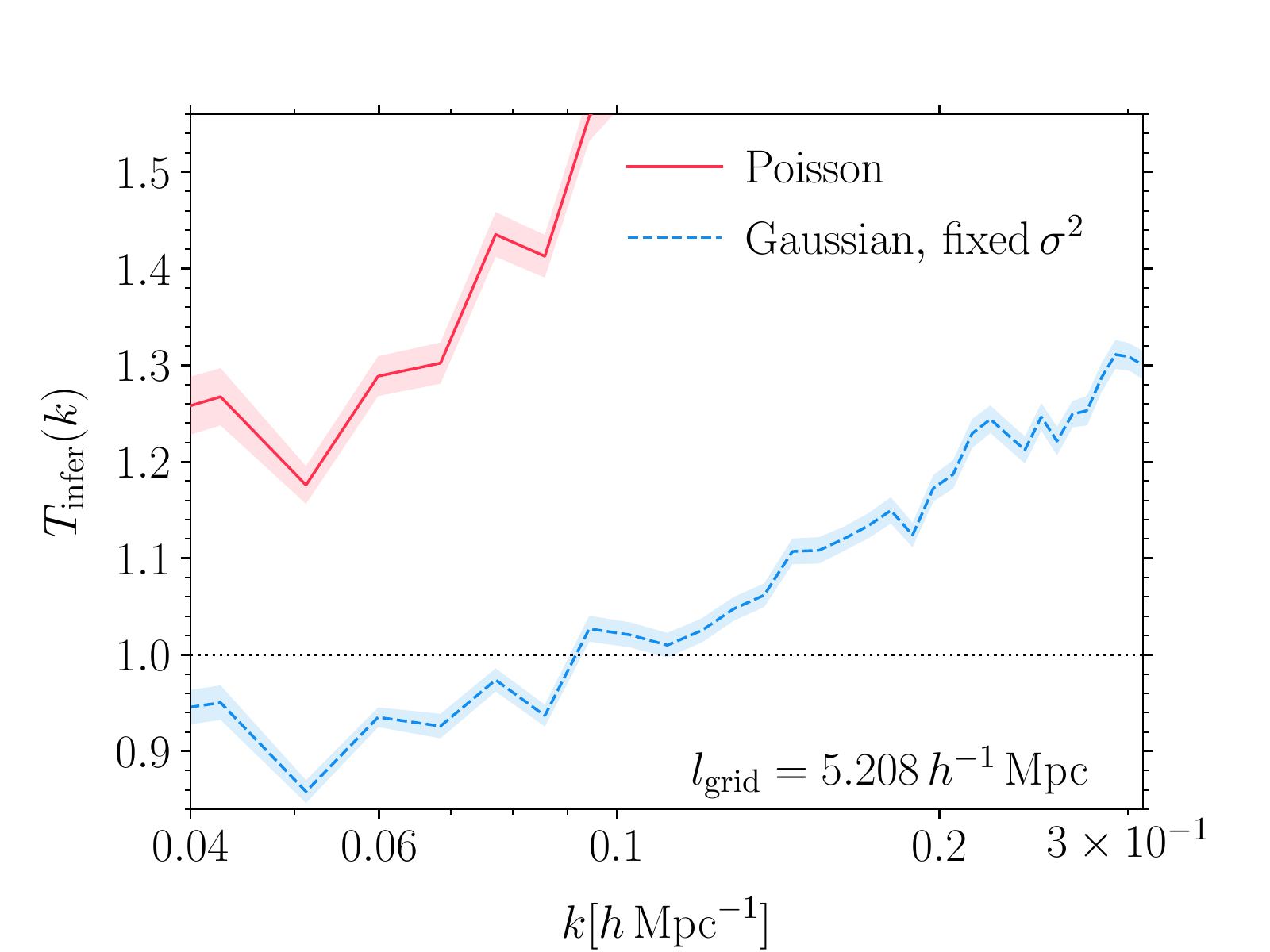}{fig:rk_ek_likelihood}{\textbf{Varying likelihood - N-body data.} \textit{Top:} The phase-correlation coefficient in different inferences with Poisson and Gaussian likelihoods as described in \refsec{likelihood}. The common Gaussian expectation is shown in dark-gray dotted line. \textit{Bottom:} The transfer function in the corresponding inferences.}
\end{figure*}

As shown in the top panels of \reffig{rk_ek_likelihood}, the phase-correlation metric marginally favors the Poisson likelihood. Interestingly, however, the transfer function strongly prefers the Gaussian case.
Perhaps it is worth keeping in mind that the amplitude-bias is robustly observed across different halo mass ranges, grid resolutions, forward models and bias models.
We therefore conclude the Poisson likelihood is \emph{incompatible} with halo stochasticity.
This is perhaps not entirely surprising as previous literature has already pointed out that halo stochasticity can be either sub-Poisson, due to halo exclusion effect \cite{Smith:2007, Hamaus:2010, Baldauf:2013}, or super-Poisson in high-density regions \cite{Miranda:2002}. Intuitively, we only expect halo stochasticity to asymptote the Poisson limit at very high wavenumber k, however, precisely in this regime, our bias expansions no longer converge due to non-linearities in the matter density field (cf. \reffig{evolved_density_histogram_L2000_N128}).

To summarize, our results demonstrate that using a Poisson likelihood to model halo stochasticity typically leads to a biased transfer function, i.e. a bias in the amplitudes of the inferred modes. This is an important implication for future cosmological inference using the field-level, forward-modeling approach. 

Note, however, that main halo and galaxy stochasticities might significantly differ: \emph{a likelihood, or a bias model for that matter, optimized on main halos might not be compatible with galaxy stochasticity or bias}.
Rather than a recommendation to always use the Gaussian rather than the Poisson likelihood, or the linear bias than a more sophisticated one, for example the broken power-law bias model, our results firmly suggest that \emph{the likelihood and the bias model should be rigorously derived from, for example, a perturbative approach, such that sub-grid systematics can be kept under control} \cite[e.g.][]{Schmidt:2019, Cabass:2019b, Cabass:2020a, Elsner:2020, Schmidt:2020a, Schmidt:2020b}.

\begin{figure*}[thbp]
	\centering
	\samelinetwofig{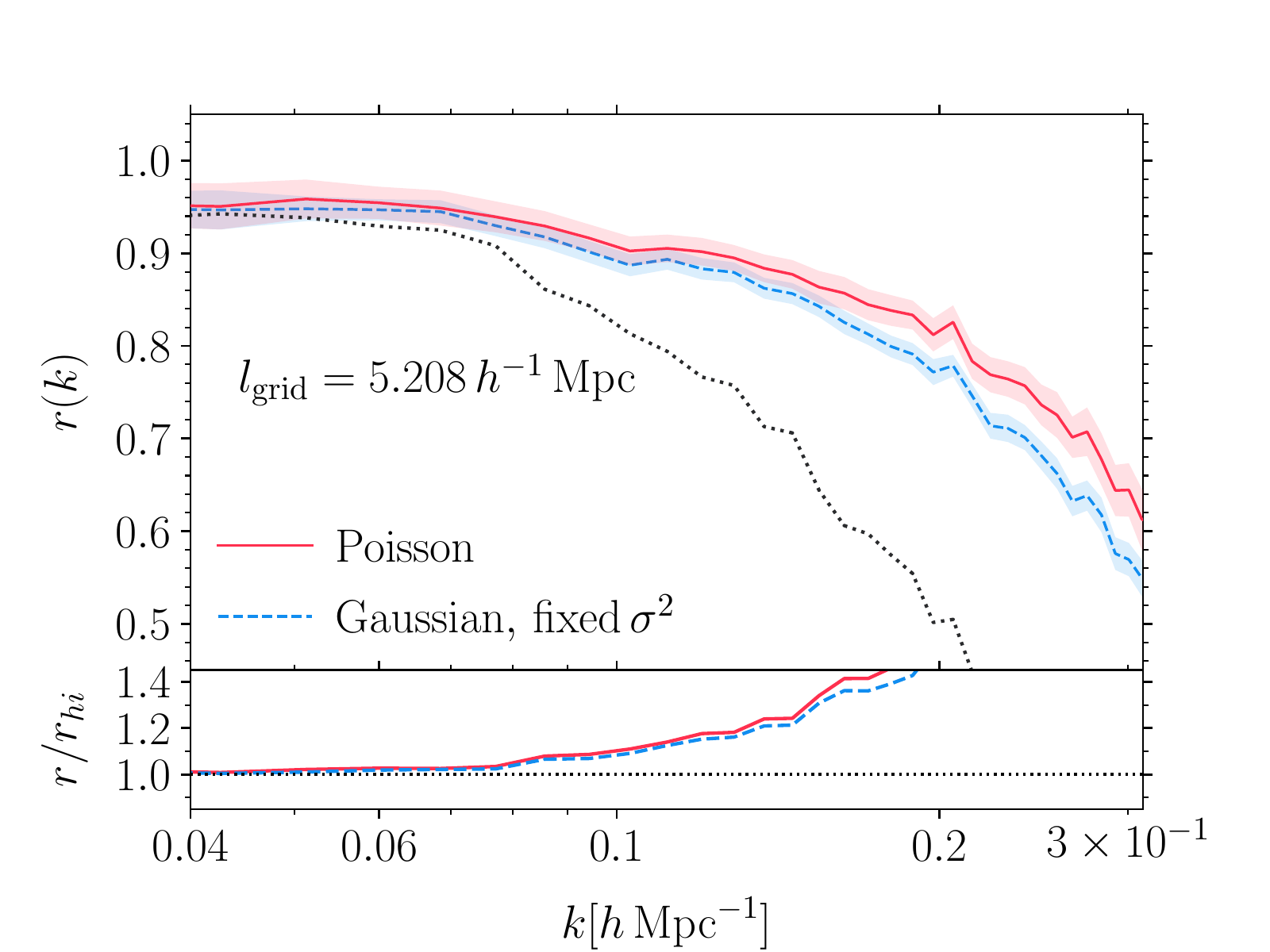}{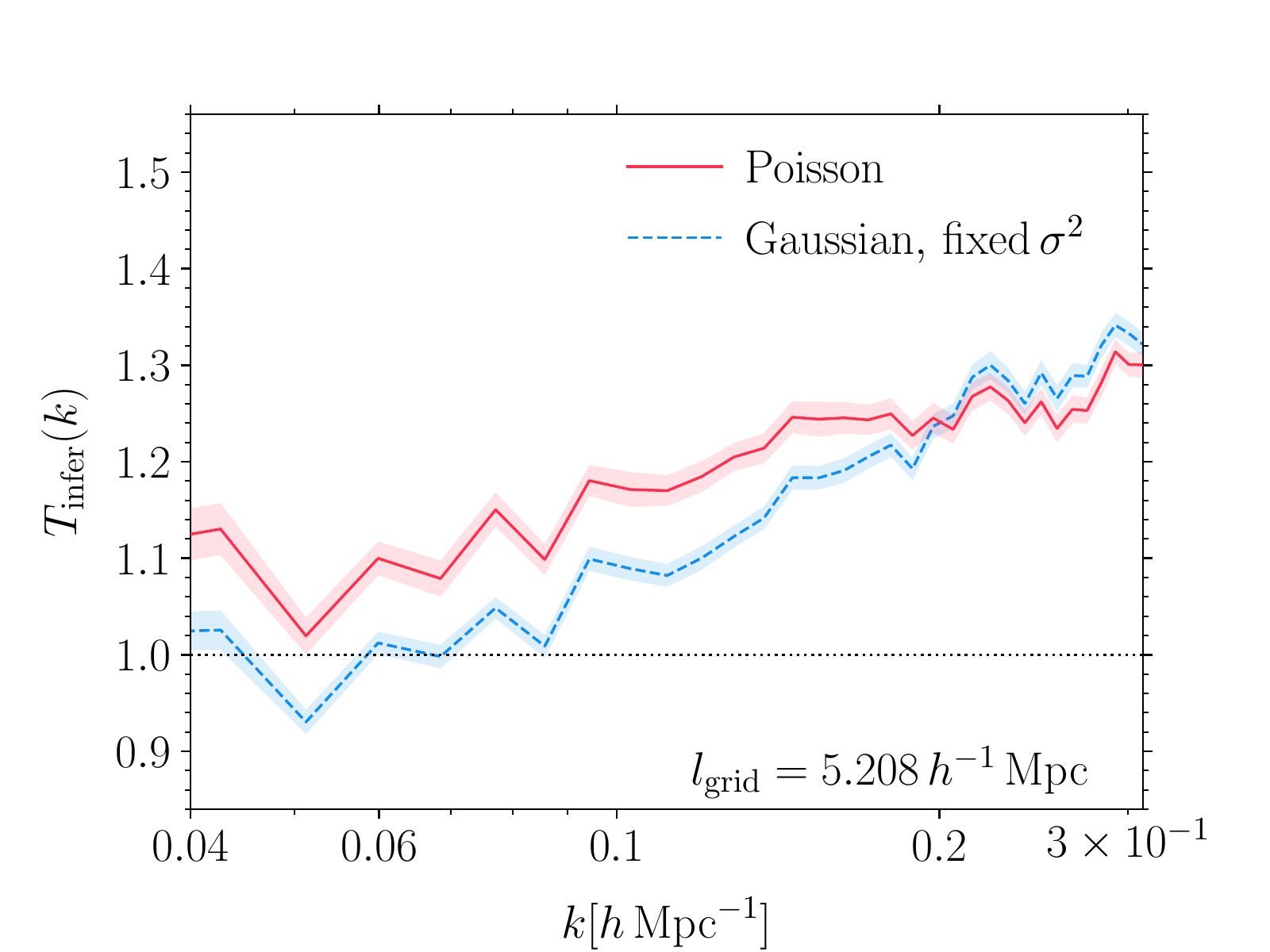}{fig:rk_ek_likelihood_linearbias}{\textbf{Revisiting the linear bias model - N-body data.}
          \textit{Left:} The phase-correlation coefficient in two inferences with Poisson and Gaussian likelihoods using the linear bias model. In the Gaussian case, we do not apply any thresholding procedure (cf. \refapp{prior_and_thresholder}) as it is unnecessary. We show the linear Gaussian expectation in dark-gray dotted line. \textit{Right:} The transfer function in the corresponding inferences.}
\end{figure*}

We now revisit the case of the linear bias model. \reffig{rk_ek_likelihood_linearbias} shows a direct comparison between the Poisson and the Gaussian likelihoods, when combined with the linear bias model, at the highest resolution we have probed. The Gaussian likelihood does not involve any thresholding procedure. Interestingly, each metric now favors a different setup: the Poisson case seems to outperform the Gaussian slightly above $k\simeq0.2\,\iMpch$ in terms of phase-correlation in the left panel of \reffig{rk_ek_likelihood_linearbias}, whereas the Gaussian case results in a less biased transfer function up to $k\simeq0.2\,\iMpch$.
 The fact that the bias in $T_{\rec}(k)$ is further reduced with the Gaussian likelihood---and without the soft-thresholder (cf. \refapp{prior_and_thresholder})---suggests that the thresholding procedure plays no significant role in our results previously shown in \reffig{rk_ek_bias_model}.
 We additionally note that the case shown in \reffig{rk_ek_likelihood_linearbias} exhibits the most significant difference in the phase correlations across all of our comparisons, going up to 5\% at $k=0.2\iMpch$.

\section{Discussion}
\label{sec:discussion}

In this paper, we have examined the forward modeling approach of biased-tracer clustering in terms of inferring the initial conditions in the observed cosmological volume. Specifically, employing the \borg framework and N-body simulations, we inspect how different ingredients in the physical data model affect the performance of the inference of the initial conditions. These include: tracer density, grid resolution, gravity model, bias model and likelihood.
We quantify the performance of each inference by two metrics:
\begin{itemize}
\item The cross correlation between inferred and input initial modes, quantified by the phase-correlation coefficient; 
\item The amplitude bias between the inferred and input amplitudes of these initial modes, quantified by the transfer function.
\end{itemize}

\subsection*{Tracer density and grid resolution} 

First, in \refsec{results_tracer_density}, using mock tracers, we observe that---in the absence of modeling error---enhancing tracer density generally improves the inference, even at scales where the clustering signal dominates over shot noise.
Similarly, increasing grid resolution increases the small-scale information the inference can access and consequently boosts the quality of the inference, as found in \refsec{results_grid_resolution}.
Both effects described above can be perfectly captured by the linear Gaussian expectation we derived in \refeq{gauss_limit_rri}.
At low $k$, the Gaussian model is an \emph{unbiased estimator} of the cross correlation between the inferred phases and the underlying truth. At high $k$, this relationship is biased but depends on neither tracer density nor grid resolution.

\subsection*{Gravity model}

Second, in \refsec{result_GADGET2}, using DM halos from N-body simulations as physical tracers, we have investigated the robustness of the inference when modeling error is present.
In particular, the choice of gravity model considered in \refsec{results_forward_model}, be it LPT and 2LPT, does not significantly affect the performance of the inference up to a moderate grid resolution of $\lgrid=15.6\,\Mpch$. This has a strong implication for cosmological inference using forward modeling approach with data from future large-volume surveys.
We note however that 2LPT outperforms LPT at higher grid resolutions.

\subsection*{Bias model and likelihood}

Last, but perhaps most interesting, we observe, for all inferences with N-body data, that the Poisson likelihood results in a significant bias on the amplitude of the inferred modes.
Further investigation is needed to determine exactly the origin of this effect, though as discussed in \refsec{results_bias_model} and \refsec{results_likelihood}, our results suggest that this is an indication of un-controlled systematics from highly non-linear fluctuations (cf. \reffig{evolved_density_histogram_L2000_N128}) and their couplings (see also results in \cite{Schmidt:2020b}, who follow an approach based on effective field theory (EFT) which adopts a sharp-$k$ filter in place of the CIC filter employed here).

Let us summarize our findings below:
\begin{itemize}
\item The phase-correlation is sensitive to a change in any ingredient. Fortunately, \emph{the shifts are generally $\lesssim5\%$ in any direction and can be well predicted by the Gaussian expectation in all cases.} This implies that analyses that cross-correlate the inference with other datasets \citep[e.g.][]{Nguyen:2020a}, will be robust to details of the physical data model.  
\item The transfer function is found to be only susceptible to the choices of bias model and likelihood. However, \emph{the differences between different likelihoods and bias models are significant.} At this point, unfortunately, it is unclear how to predict these from first principles.
\end{itemize}

\section{Conclusions}
\label{sec:conclusion}

To our knowledge, our results represent the most stringent test for Bayesian forward modeling and inference of biased tracer clustering, within the \borg framework, so far. Our input data are main halos identified in N-body simulations, unlike the mock tracers generated from the same physical data model employed by previous analyses \citep[see, e.g.][]{Jasche:2013, Ramanah:2019}.
More realistic and rigorous tests are still awaiting ahead in order to achieve the goal of unbiased inference of initial conditions and cosmological parameters using this approach. Specifically, extending the input biased tracers to include galaxies---either populated inside DM halos using some halo occupation distribution model, or directly identified in hydrodynamic simulations---would be such a test. An even more realistic case must also consider RSD, survey geometries and selection effects applied to the galaxy field. We defer such studies to future work as well.

Clearly, a transfer function $T_{\rec}(k)$ with a scale-dependent bias up to $\sim30\%$ has a significant implication for cosmological inference, especially when future analyses are aiming to reach percent-level constraints on cosmological parameters. In particular, one would expect biases in not only $\A_s$ but also $\Om$, $\ns$ and probably the distance scale as well. We thus view this issue as a call for attention to studies and developments of accurate bias models and likelihoods. As already pointed out above, \borg and the work in this paper present a physically-motivated and technically-ready statistical framework for such studies.

Specifically, an EFT-based approach such as the one presented in \cite{Schmidt:2019, Elsner:2020, Schmidt:2020b} is certainly preferable. In fact, \cite{Schmidt:2020b} showed that for more conservative scale cut-offs, one can recover an unbiased primordial power spectrum normalization within percent-level for the same fiducial halo sample studied in this work. The analysis in \cite{Schmidt:2020b} however fixed the phases. It would be interesting to investigate how the EFT approach performs, specifically in terms of the level of systematic bias, when the phases are inferred from the data and marginalized over, as done in this paper. We leave this investigation for an upcoming work.

\FloatBarrier

   \acknowledgments

   We would like to thank Eiichiro Komatsu, Alex Barreira, Giovani Cabass, Ariel Sanchez as well as all Aquila Consortium members for helpful discussions.
   We especially thank Titouan Lazeyras for sharing the access to the N-body simulations.
   We extend our thanks to Elisa Ferreira and Giovanni Aric\`{o} for useful comments that helped improving the plots.
   MN and FS acknowledge support from the Starting Grant (ERC-2015-STG 678652) \enquote{GrInflaGal} of the European Research Council.
   GL acknowledges financial support from the ILP LABEX (under reference ANR-10-LABX-63) which is financed by French state funds managed by the ANR within the Investissements d'Avenir programme under reference ANR-11-IDEX-0004-02. This work was supported by the ANR BIG4 project, grant ANR-16-CE23-0002 of the French Agence Nationale de la Recherche.
   This work is done within the Aquila Consortium.\footnote{\url{https://aquila-consortium.org}}

\clearpage

\appendix

\section{Gaussian expectation of the Fourier phase-correlation}
\label{app:Gaussian_limit_of_rk}

In this appendix we derive the Gaussian expectation of the phase-correlation coefficient, as shown in \refeq{gauss_limit_rri}.

As discussed in \refsec{results}, within this limit, we can assume that all fields (including the noise field) are Gaussian, and assume a \emph{linear} bias relation of the following form:
\be
\dh(\vk) = b_1\d_{m,\lin}^{\mathrm{input}}(\vk) + \varepsilon_h(\vk),
\label{eq:gauss_limit_linear_bias}
\ee
where $\d_{m,\lin}^{\mathrm{input}}$ here denotes the input matter density field \emph{linearly extrapolated} to redshift zero.
Since the halo stochasticity $\varepsilon_h$ has zero mean and is assumed to be uncorrelated with the matter density field, the optimal estimator for the reconstructed density field is simply given by
\be
\d_{m,\lin}^{\rec}(\vk) = \frac1{b_1} \dh(\vk).
\label{eq:gauss_limit_rec}
\ee
This together with \refeq{gauss_limit_linear_bias} implies that 
\be
\<\d_{m,\lin}^{\rec}(\vk)\d_{m,\lin}^{\mathrm{input}}(\vk)\>_k =
\<\d_{m,\lin}^{\mathrm{input}}(\vk)\d_{m,\lin}^{\mathrm{input}}(\vk)\>_k = \Plin(k),
\label{eq:gauss_limit_rec_in}
\ee
where $\Plin$ denotes the linear matter power spectrum at $z=0$.
In addition, the power spectrum of the inferred modes can be expressed as (cf. \refeq{gauss_limit_rec} and \refeq{gauss_limit_linear_bias})
\bab
\<\d_{m,\lin}^{\rec}(\vk)\d_{m,\lin}^{\rec}(\vk)\>_k &= \frac{1}{b_1^2} \<\dh(\vk)\dh(\vk)\>_k \\
&= \Plin(k) + \frac1{b_1^2} P_\varepsilon(k) .
\label{eq:gauss_limit_rec_rec}
\eab
Combining \refeq{gauss_limit_rec_in} with \refeq{gauss_limit_rec_rec} then
yields the desired cross-correlation coefficient between $\d_{m,\lin}^{\rec}$ and $\d_{m,\lin}^{\rm input}$:
\be
r_{\mathrm{ri},\lin}(k) = \frac{1}{\sqrt{1+P_\varepsilon(k)/\left(b_1^2\Plin(k)\right)}}.
\label{eq:gauss_limit_rri_app}
\ee

\section{Residual bias due to normalization using $\se$}
\label{app:normalization_bias}

\begin{figure*}[thbp]
\centerline{\resizebox{0.6\textwidth}{!}{
	\includegraphics*{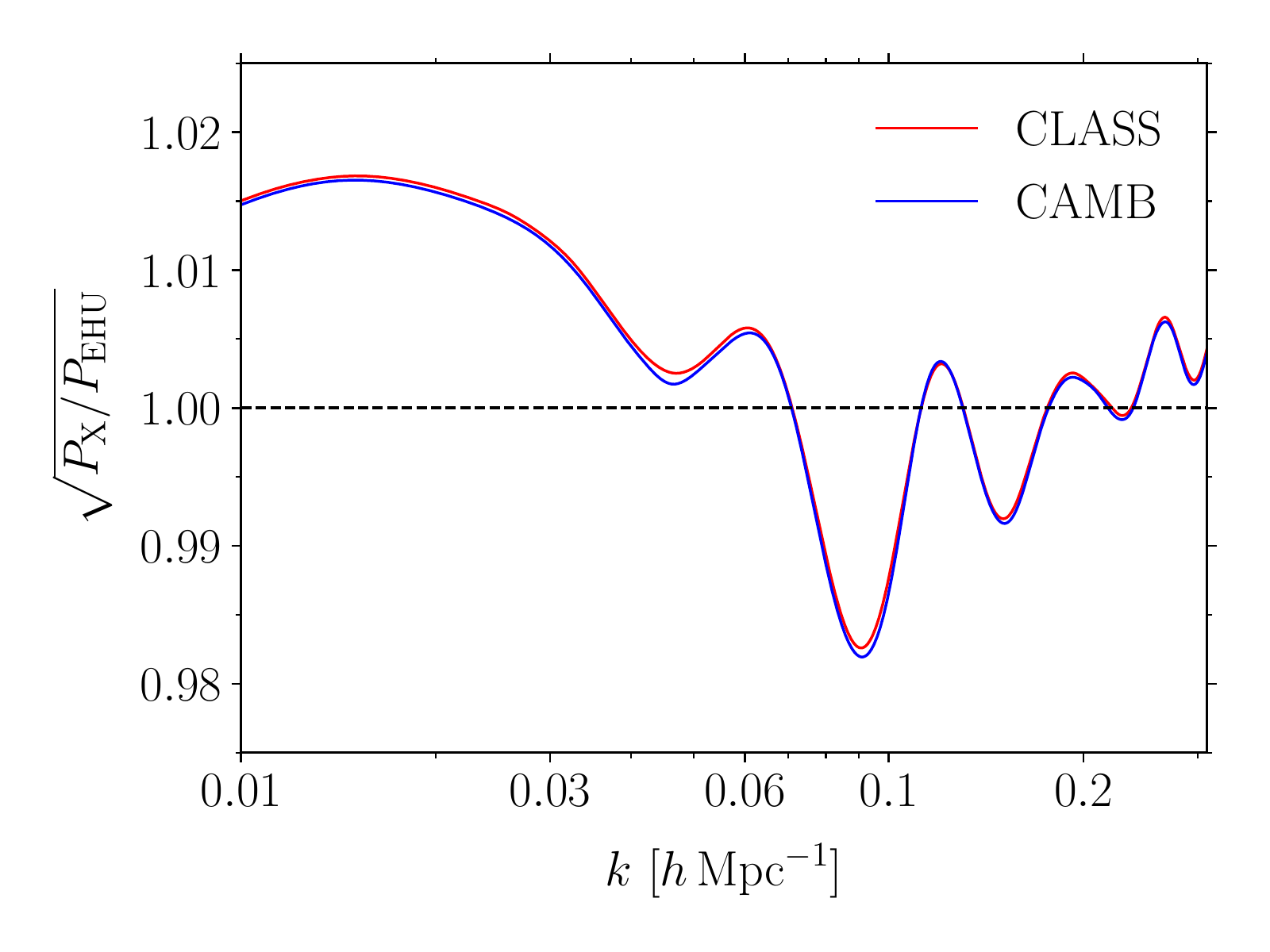}}}\caption{Ratios between the transfer functions at redshift zero, computed by CLASS (red) or CAMB (blue) and that determined from Eisenstein-Hu fitting formula.}	\label{fig:transfer_function_bias_due_to_normalization}
\end{figure*}

Our N-body simulations use input matter transfer functions computed by Boltzmann codes CLASS and CAMB, respectively. On the other hand, for performance reasons, \borg employs  
the physically-motivated Eisenstein-Hu fitting function \cite{eisenstein/hu:1998,eisenstein/hu:1999} to compute the same quantity. In either cases, the matter transfer functions were outputted at redshift zero, then back-scaled to the initial redshifts in the simulation and the forward model. Further, \borg cosmological setting requests one to specify the normalization of the \emph{linear matter power spectrum at redshift zero} $\se$ instead of the normalization of the \emph{primordial power spectrum} $\A_s$.
This convention, plus the difference between matter transfer functions computed by CLASS or CAMB and those obtained from the Eisentein-Hu fitting formula, together will result in a bias between the inferred and input initial conditions, and hence a deviation of $T_{\rm rec}(k)$ from 1. In this appendix, we examine this effect by measuring the ratio between these transfer functions. Specifically, \reffig{transfer_function_bias_due_to_normalization} shows that one can expect a \emph{scale-dependent} bias in $T_{\rm rec}$ up to 1.7\% on the scales of interest in this paper.
This clearly cannot explain the bias observed in \refsec{result_GADGET2}.

 \section{Priors and thresholding}
\label{app:prior_and_thresholder}

For all inferences presented in this paper, we impose uniform priors and positivity constraints on all bias parameters, including the nuisance parameter $n_0$:
\be
\P_{\mathrm{prior}}\left(n_0,\{b_O\}\right) = \Theta(n_0)\prod_O\Theta(b_O),
\ee
in which $\Theta(x)$ is the Heaviside step function.

The Poisson likelihood in \refeq{cell_Poisson_likelihood} requires a non-negative $\nhdet$; in order to satisfy this condition in the particular case of the linear bias model, we additionally implement a soft-thresholder of the form:
\be
\nhdet = n_0\begin{cases}
\left(1+\dhdet\right) &\text{$\dhdet>-0.9$} \\
th\left(1+\dhdet, c\right) &\text{otherwise},
\end{cases}
\label{eq:smooth_boundary}
\ee
with
\be
th\left(1+\dhdet, c\right) = \ln \left(1+\exp\left[c\left(1+\dhdet\right)\right]\right)/c
\label{eq:thresholder}
\ee
being the softplus function \cite{Glorot:2011}. We choose $c=5$ for our implementation.
We emphasize again that this thresholding procedure is only needed for, and thus applied to, inferences using both linear bias model and Poisson likelihood.

\section{MCMC burn-in phase and chain thinning}
\label{app:burnin_and_thinning}
This appendix describes in detail how we verify that our MCMC chains have indeed approached the target typical sets and how we select MCMC samples in our analysis.

All the Markov chains considered in this paper were initialized in an over-dispersed state, presumably remotely far from the stationary distribution region in the posterior space. We refer to this initial phase as the \emph{warm-up} phase.
For each chain, we ensure that samples included in our analysis do not come from this phase by carefully tracing the posterior power spectrum and bias parameters.
In \reffig{trace_plot_examples}, we show two such examples. Both trace plots confirm that the chain has reached a stationary distribution after around 2000 samples.

To achieve a reasonable effective sample size while avoiding wasting disk storage, we only store and analyze 1-in-every-20 MCMC sample.

\begin{figure*}[thbp]
	\centering
	\samelinetwofig{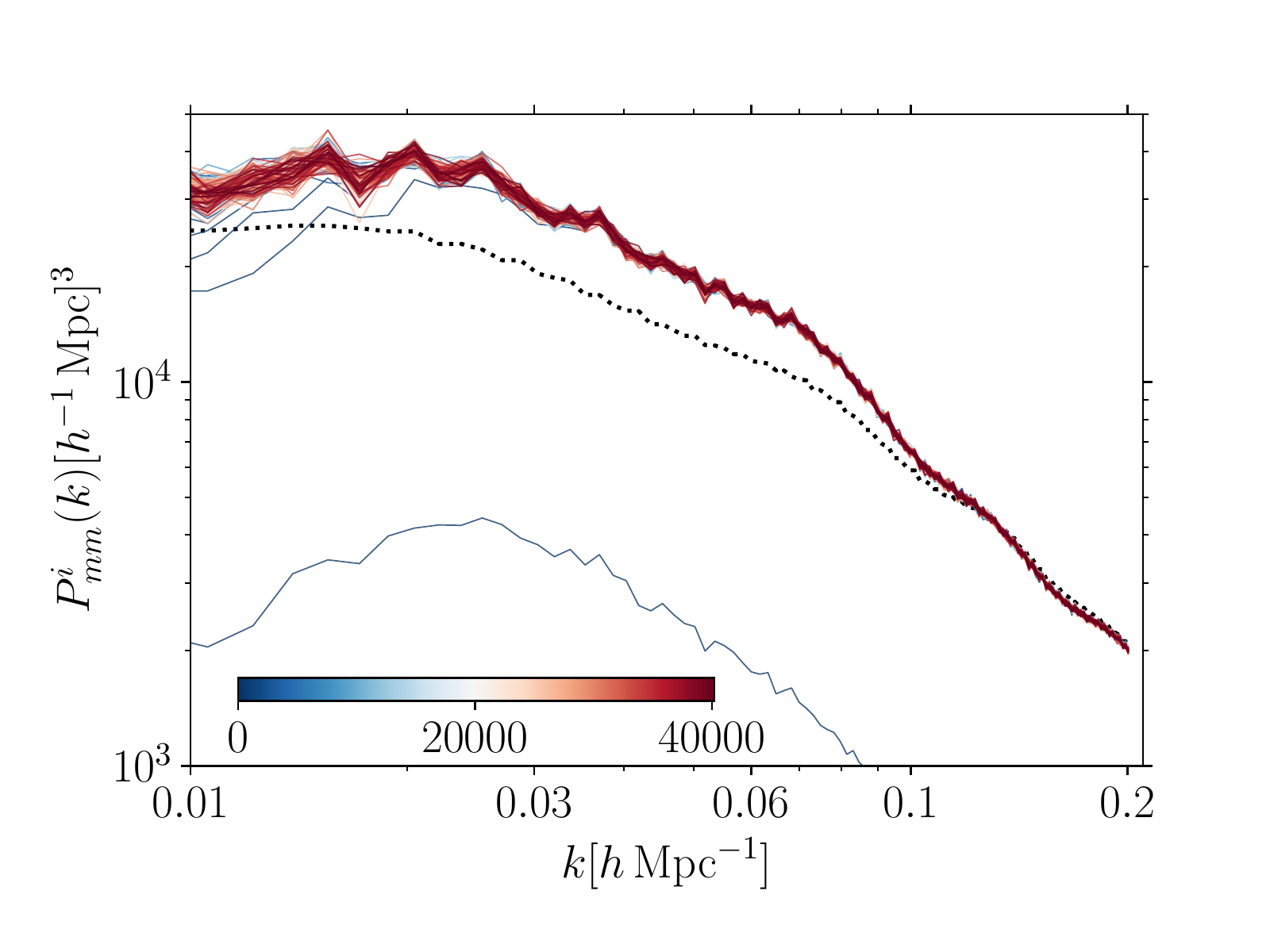}{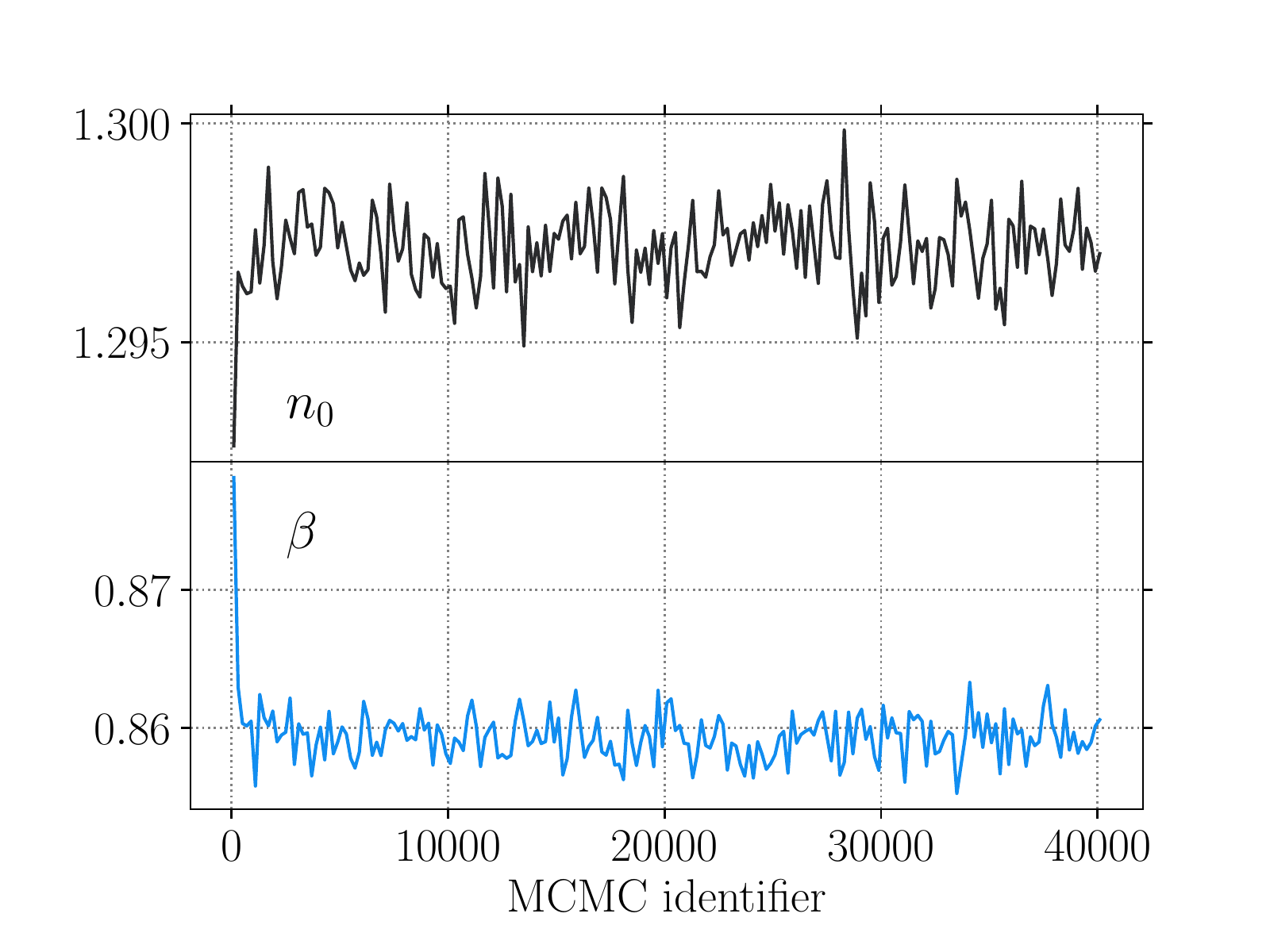}{fig:trace_plot_examples}{\textit{Left:} One example of tracing the power spectrum posterior during and after warm-up phase. For visual clarity , we only plot 1-in-every-200 MCMC samples. The color scale indicates the MCMC identifier. The dotted line denotes the input power spectrum for the given cosmology. The low-k bias observed here corresponds to the bias in the transfer function seen in the bottom left panel of \reffig{rk_ek_tracer_density_grid_resolution_halo} (red line there). \textit{Right:} Trace plot of the bias parameters in the same inference. Again, each point in this plot represents 1-in-every-200 MCMC samples. Further, samples whose identifier is smaller than 100 were removed to enhance visibility of the y-axis.}
\end{figure*}
 
 \clearpage

\bibliographystyle{JHEP}
\bibliography{references}
\end{document}